\documentclass[fleqn,usenatbib]{mnras}

\usepackage{newtxtext,newtxmath}

\usepackage[T1]{fontenc}

\DeclareRobustCommand{\VAN}[3]{#2}
\let\VANthebibliography\thebibliography
\def\thebibliography{\DeclareRobustCommand{\VAN}[3]{##3}\VANthebibliography}

\usepackage{graphicx}	
\usepackage{amsmath}	

\usepackage{xcolor}
\usepackage{siunitx}
\usepackage{layouts}

\usepackage{booktabs}
\usepackage[export]{adjustbox}

\DeclareSIUnit[]{\year}{yr}
\DeclareSIUnit[]{\parsec}{pc}
\DeclareSIUnit[]{\dex}{dex}

\title[Evolution of the thin disc radial $\mathrm{[Fe/H]}$ gradient]{The evolution of the Milky Way's thin disc radial metallicity gradient with K2 asteroseismic ages}

\author[E. Willett et al.]{%
Emma Willett,$^{1}$\thanks{E-mail: EMW575@student.bham.ac.uk}
Andrea Miglio,$^{2, 3, 1}$
J. Ted Mackereth,$^{4, 5}$
Cristina Chiappini,$^{6}$
Alexander J. Lyttle,$^{1}$
\newauthor
Yvonne Elsworth,$^{1}$
Benoît Mosser,$^{7}$
Saniya Khan,$^{8}$
Friedrich Anders,$^{9, 10, 11}$
Giada Casali$^{2, 3}$ and
Valeria Grisoni$^{2, 3}$
\\
$^{1}$ School of Physics and Astronomy, University of Birmingham, Edgbaston, Birmingham, B15 2TT, UK\\
$^{2}$ Dipartimento di Fisica e Astronomia, Università degli Studi di Bologna, Via Gobetti 93/2, I-40129 Bologna, Italy\\
$^{3}$ INAF - Osservatorio di Astrofisica e Scienza dello Spazio di Bologna, Via Gobetti 93/3, I-40129 Bologna, Italy\\
$^{4}$ Canadian Institute for Theoretical Astrophysics, University of Toronto, Toronto, ON M5S 3H8, Canada\\
$^{5}$ Dunlap Institute for Astronomy and Astrophysics, University of Toronto, Toronto, ON M5S 3H4, Canada\\
$^{6}$ Leibniz-Institut fur Astrophysik Potsdam (AIP), An der Sternwarte 16, D-14482 Potsdam, Germany\\
$^{7}$ LESIA, Observatoire de Paris, Universit\'e PSL, CNRS, Sorbonne Universit\'e, Universit\'e de Paris, 92195 Meudon, France\\
$^{8}$ Institute of Physics, Laboratory of Astrophysics, \'Ecole Polytechnique F\'ed\'erale de Lausanne (EPFL), Observatoire de Sauverny, 1290 Versoix, Switzerland\\
$^{9}$ Departament de Física Quàntica i Astrofísica (FQA), Universitat de Barcelona (UB),  Martí i Franquès, 1, 08028 Barcelona, Spain\\
$^{10}$ Institut de Ciències del Cosmos (ICCUB), Universitat de Barcelona (UB), Martí i Franquès, 1, 08028 Barcelona, Spain\\
$^{11}$ Institut d'Estudis Espacials de Catalunya (IEEC), Gran Capità, 2-4, 08034 Barcelona, Spain
}

\date{Accepted XXX. Received YYY; in original form ZZZ}

\pubyear{2022}

\begin{document}
\label{firstpage}
\pagerange{\pageref{firstpage}--\pageref{lastpage}}
\maketitle

\begin{abstract}
The radial metallicity distribution of the Milky Way's disc is an important observational constraint for models of the formation and evolution of our Galaxy. It informs our understanding of the chemical enrichment of the Galactic disc and the dynamical processes therein, particularly radial migration. We investigate how the metallicity changes with guiding radius in the thin disc using a sample of red-giant stars with robust astrometric, spectroscopic and asteroseismic parameters. Our sample contains $668$ stars with guiding radii $\qty[]{4}{\kilo\parsec} < R_\mathrm{g} < \qty[]{11}{\kilo\parsec}$ and asteroseismic ages covering the whole history of the thin disc with precision $\approx 25\%$. We use MCMC analysis to measure the gradient and its intrinsic spread in bins of age and construct a hierarchical Bayesian model to investigate the evolution of these parameters independently of the bins. We find a smooth evolution of the gradient from $\approx \qty[per-mode=symbol]{-0.07}{\dex\per\kilo\parsec}$ in the youngest stars to $\approx \qty[per-mode=symbol]{-0.04}{\dex\per\kilo\parsec}$ in stars older than $\qty[]{10}{\giga\year}$, with no break at intermediate ages. Our results are consistent with those based on asteroseismic ages from CoRoT, with that found in Cepheid variables for stars younger than $\qty[]{1}{\giga\year}{}$, and with open clusters for stars younger than $\qty[]{6}{\giga\year}{}$. For older stars we find a significantly lower metallicity in our sample than in the clusters, suggesting a survival bias favouring more metal-rich clusters. We also find that the chemical evolution model of \citet{2009IAUS..254..191C} is too metal-poor in the early stages of disc formation. Our results provide strong new constraints for the growth and enrichment of the thin disc and radial migration, which will facilitate new tests of model conditions and physics.
\end{abstract}

\begin{keywords}
asteroseismology -- Galaxy: abundances -- Galaxy: disc -- Galaxy: evolution -- Galaxy: stellar content -- stars: abundances
\end{keywords}




\section{Introduction}
\label{sec:Introduction}

Over the last decade, the field of Galactic archaeology has expanded rapidly, with many new discoveries concerning the formation and evolution of the Milky Way (MW). These advances have been facilitated by the availability of astrometry of unprecedented precision from \textit{Gaia} \citep{2016A&A...595A...1G} and several large spectroscopic surveys, such as the Apache Point Observatory Galactic Evolution Experiment \citep[APOGEE;][]{2017AJ....154...94M}, \textit{Gaia}-ESO \citep{2022A&A...666A.121R}, GALactic Archeology with HERMES \citep[GALAH;][]{2021MNRAS.506..150B}, the Large sky Area Multi-Object fibre Spectroscopic Telescope \citep[LAMOST;][]{2012RAA....12..735D, 2012arXiv1206.3569Z} and the RAdial Velocity Experiment \citep[RAVE;][]{2020AJ....160...82S, 2020AJ....160...83S}. When combined with stellar ages, these provide meaningful tests of chemo-dynamical models of the MW.

Many such models of spiral galaxies predict that the inner part of the disc forms earlier than the outer regions \citep[e.g.][]{1976MNRAS.176...31L, 1989MNRAS.239..885M, 1997ApJ...477..765C, 2019ApJ...884...99F}, and the idea of `inside-out' formation is now widely accepted. One of the signatures of this formation scenario is a radial metallicity\footnote{Throughout this work we use iron abundance as a tracer for metallicity: $\mathrm{\left[Fe/H\right]} = \log_{10}\left(N_\mathrm{Fe}/N_\mathrm{H}\right)_* - \log_{10}\left(N_\mathrm{Fe}/N_\mathrm{H}\right)_\odot$, where $N_\mathrm{Fe}$ and $N_\mathrm{H}$ are the numbers of iron and hydrogen nuclei per unit volume of the stellar photosphere, respectively.} gradient in the Galactic disc, with higher metallicities in the centre, caused by more generations of stars which enrich the material from which the next generation form. This gradient might change with time as the disc expands and star formation proceeds unevenly. In the MW we have the opportunity to explore this evolution by using tracers of the metallicity gradient with different ages, in contrast to studies of other galaxies where, from a given tracer, we see only a snapshot of the aggregate effect \citep[e.g.][]{2010Natur.467..811C, 2016A&A...588A..91M}.

In addition to its intrinsic interest as a property of the MW, the radial metallicity gradient also helps to constrain dynamical processes in the disc. If we expect that stars are born into the Galaxy with some tight relationship between their metallicity and birth radius, $R_\mathrm{b}$, but the relationship in a mono-age population shows some scatter, then we may infer that some stars have moved away from $R_\mathrm{b}$. This assumption breaks down in the presence of contaminants from the debris of galactic mergers of other processes, but it is an acceptable simplification in a robustly selected in-situ population. Inferring birth radii is an important step towards recovering the MW's history, and therefore tight observational constraints on the metallicity gradient are very valuable \citep[e.g.][]{1996A&A...314..438W, 2013A&A...558A...9M, 2018MNRAS.481.1645M, 2018ApJ...865...96F}.

There are two reasons why a star may be observed with a galactocentric radius, $R_\mathrm{Gal} \neq R_\mathrm{b}$. First, the eccentricity of the orbit naturally results in a star being observed at different $R_\mathrm{Gal}$ during the orbit. Stars in the disc are born onto nearly circular orbits, which increase in eccentricity over time as a result of radial heating. This process increases the radial action, $J_R$, and is referred to as `blurring' \citep{2002MNRAS.336..785S, 2009MNRAS.396..203S}. Importantly, for efforts to recover birth radii, blurring does not change the orbital angular momentum, $L_z$, so the star retains some `kinematic memory' of its birth conditions. By considering the metallicity gradient with guiding radius, $R_\mathrm{g}$ (the radius of a circular orbit of the same angular momentum) we can mitigate this effect in our sample.

Second, interactions with non-axisymmetric features of the MW's disc (for example, the bar or spiral arms) can cause `churning' \citep{2002MNRAS.336..785S}, in which an exchange of angular momentum shifts the star into a new orbit. \citet{2020ApJ...896...15F} found that this effect is roughly an order of magnitude stronger than blurring in the Galactic disc, shifting stars by up to $\approx \qty[]{3}{\kilo\parsec}$ over \qty[]{6}{\giga\year} on average. The change of $L_z$ means that $R_\mathrm{g} \neq R_\mathrm{b}$ and $R_\mathrm{b}$ can only be inferred by indirect means, such as the combination of the metallicity gradient and age. \citet{2016AN....337..944M} found that churning is necessary in their chemo-dynamical simulation in order to reproduce the observed spread of metallicity in stars of a given age. The same model also shows that the presence of churning affects the flaring of a galactic disc in a quiescent setting or during a merger \citep{2016AN....337..703M}.

Constraints on migratory processes like churning require precise measurements of the metallicity gradient in populations spanning the full history of the disc. This necessitates a large sample, covering a wide range of radii, metallicities and ages. Open clusters (OCs) have been used in such studies, \citep{2009A&A...494...95M, 2017MNRAS.470.4363C, 2019MNRAS.490.1821C, 2021MNRAS.503.3279S, 2022Univ....8...87S, 2022arXiv220605534G, 2022AJ....164...85M, 2022MNRAS.509..421N}, however, dynamical processes are thought to create a survival bias towards high-metallicity OCs, obscuring the underlying chemical evolution of the Galaxy \citep{1987A&A...188...35L, 2017A&A...600A..70A, 2021MNRAS.503.3279S}. Cepheid variables can be used to probe the present-day gradient, as they provide an excellent radial range for targets with $\qty[]{20}{\mega\year} < \tau < \qty[]{400}{\mega\year}$ \citep{2014A&A...566A..37G}. Other young tracers such as OB stars \citep{1994A&A...289..740K, 2004ApJ...617.1115D, 2019A&A...625A.120B} and HII regions \citep{2011ApJ...738...27B, 2005ApJ...618L..95E} can be used to trace the [O/H] gradient over large regions of the Galaxy but, as for Cepheids, they cannot reveal the evolution over its history.

Low-mass field stars are ideal tracers, but there are many well-documented challenges associated with obtaining reliable stellar ages for large samples. Ages may be estimated by comparison to isochrones, but they are sensitive to changes in model physics or temperature scale and are generally restricted to lower-luminosity phases of stellar evolution. For red giants which exhibit solar-like oscillations, asteroseismic analysis may be used to obtain precise and robust (though not model-independent) stellar ages \citep{2010ARA&A..48..581S, 2013ARA&A..51..353C, 2014EAS....65..177L}. \citet{2017A&A...600A..70A} demonstrated the potential of this approach with the CoRoGEE sample \citep{2017A&A...597A..30A}, combining asteroseismic constraints from two fields observed by CoRoT \citep{2006cosp...36.3749B} with spectroscopy from APOGEE. The combination of asteroseismology and spectroscopy offers many advantages over classical tracers, as we can probe the full history of the disc over several \qty[]{}{\kilo\parsec}.

\textit{Kepler} \citep{2010PASP..122..131G} represents the ‘gold standard’ in asteroseismic data, enabling ages to be determined to 10\% precision \citep{2021NatAs...5..640M}. However, while the signature of radial migration can be seen in \textit{Kepler} data (e.g. Figure 6 of \citet{2021A&A...645A..85M}), they are restricted in spatial coverage to a single \qty[]{105}{\deg\squared}{} area. In this work we utilise the results of the successor mission, K2 \citep{2014PASP..126..398H}, which provides observations from all around the ecliptic, split into 20 campaigns each observed for around \qty[]{80}{days}. Though the observation time per campaign was much shorter in duration than \textit{Kepler} for a single field, the global oscillation frequencies are sufficient to place good constraints on the stellar mass, $M_*$, with which the age of low-mass red giants is tightly correlated. Combining these data with chemical abundances from APOGEE DR17 \citep{2022ApJS..259...35A}, and \textit{Gaia} DR3 \citep{2022arXiv220800211G} unlocks the potential of asteroseismic ages for Galactic archaeology on a large scale.

In this work, we compile a catalogue of K2 targets with robust and homogeneously determined stellar parameters from asteroseismology. From this, we select members of the thin disc and measure the radial metallicity gradient as a function of age with two complementary techniques. We introduce our sample in Section \ref{sec:Observations} and describe our analysis in Sections \ref{sec:Bins} and \ref{sec:HBM}. Section \ref{sec:Discussion} contains a detailed discussion of our results and a comparison to the existing literature and theoretical predictions. Finally, our conclusions are presented in Section \ref{sec:Conclusions}.

\section{Observational constraints}
\label{sec:Observations}

In this section, we describe our sample of K2 solar-like oscillating red giants (Section \ref{ssec:Sample_obs}), outline how the stellar and orbital parameters were inferred from the observables (Section \ref{ssec:Sample_inf}), and how members of the thin disc were selected, utilising both the abundance and kinematic information (Section \ref{ssec:DiscSel}).

\subsection{Asteroseismic, Spectroscopic and Astrometric Constraints}
\label{ssec:Sample_obs}

Our sample is based upon the catalogue of Willett et al. (in preparation), which contains asteroseismic parameters from the pipelines of \citet{2009A&A...508..877M} and \citet{2020RNAAS...4..177E}, as well as spectroscopically derived abundances and radial velocities (RVs) from APOGEE DR17 \citep{2022ApJS..259...35A}, and 5-parameter astrometric solutions from \textit{Gaia} DR3 \citep{2022arXiv220800211G}.

In order to obtain a set of homogeneously constrained stellar ages, in this work we consider global asteroseismic parameters $\nu_\mathrm{max}$, the frequency of maximum oscillation power, and $\Delta\nu$, the large frequency separation, from the pipeline of \citet{2020RNAAS...4..177E} only. This pipeline has been shown to produce fewer false positive detections \citep{2021MNRAS.502.1947M} and the definition of $\Delta\nu$ (which varies between pipelines) is more similar to that used in the stellar parameter estimation code PARAM. The resulting sample is spread across K2 campaigns 1 -- 8 and 10 -- 18 which provides good coverage out to $\approx$ \qty[]{3}{kpc} from the solar position. 

The abundances used in this work are those produced by the APOGEE Stellar Parameters and Chemical Abundances Pipeline \citep[ASPCAP;][]{2016AJ....151..144G}. A full description of this pipeline as applied to APOGEE DR17 is given in Holtzman et al. (in preparation). We remove targets with \texttt{ASPCAPFLAG} containing \texttt{STAR\_BAD} or \texttt{STAR\_WARN}, and those with any \texttt{RV\_FLAG} set. 

We use five-parameter astrometric solutions [$\alpha, \delta, \varpi, \mu_\alpha, \mu_\delta$] from \textit{Gaia} DR3. Following \citet{LL:LL-124}, we remove targets with \texttt{ruwe} $> 1.4$ or which are marked as binaries by the \texttt{non\_single\_star} flag. 

The final sample comprises 5885 targets, shown in Figure \ref{fig:XY_RZ}, with median metallicity uncertainty $\sigma_\mathrm{[Fe/H]} = \qty[]{0.008}{\dex}$. In practice, we enforce a lower limit of $\sigma_\mathrm{[Fe/H]} > \qty[]{0.05}{\dex}$ and $\sigma_{T_\mathrm{eff}} > \qty[]{50}{\kelvin}$ during the inference of the stellar parameters (Section \ref{ssec:Sample_inf}) because the pipeline uncertainties reflect only the internal errors. In addition, inflating the error on $T_\mathrm{eff}$ helps to mitigate the uncertainties associated with the descriptions of near-surface convection and outer boundary conditions in the stellar models. We use the ASPCAP $\sigma_\mathrm{[Fe/H]}$ in our analysis and the metallicity gradient, but show the negligible effect on our results of inflating the [Fe/H] uncertainty in Appendix \ref{app:feherr}.

\subsection{Inferred Stellar and Orbital Parameters}
\label{ssec:Sample_inf}

In this work we use the stellar parameters described in Willett et al. (in preparation), specifically the mass, $M_*$, age, $\tau$, and distance, $D$. These parameters are inferred using the PARAM code \citep{2006A&A...458..609D, 2014MNRAS.445.2758R, 2017MNRAS.467.1433R}, which performs a Bayesian comparison to a grid of stellar models and returns full posterior information. Asteroseismic constraints are provided by $\Delta\nu$ and $\nu_\mathrm{max}$ and we use the reference model grid (labelled G2) from the work of \citet{2021A&A...645A..85M}. We remove stars with $\nu_\mathrm{max}$ more than three standard deviations below \qty[]{20}{\micro\hertz}{}, as asteroseismic inferences have not been extensively tested in this domain. A detailed discussion of the asteroseismic distances and how they compare to those from \textit{Gaia} can be found in \citet{2023arXiv230407158K}.

Orbital parameters are obtained from the same source and here we use the guiding radius, $R_\mathrm{g}$, and maximum excursion from the Galactic plane $Z_\mathrm{max}$. They are computed using the fast orbit estimation method of \citet{2018PASP..130k4501M}, implemented in \texttt{galpy} \citep{2015ApJS..216...29B}. Throughout this work, we adopt the simple Milky Way potential \texttt{MWPotential2014} \citep{2015ApJS..216...29B} and, following this model, we assume the radial position of the Sun to be $R_{\mathrm{Gal},\,\odot} = \qty[]{8.0}{\kilo\parsec}$, where the circular velocity is $v_\mathrm{circ} = \qty[]{220}{\kilo\meter\per\second}$ \citep{2012ApJ...759..131B}. The additional solar motion is given by $[U, V, W]_\odot = [-11.1, 12.24, 7.25]\qty[]{}{\kilo\meter\per\second}$ \citep{2010MNRAS.403.1829S}, and the Sun's vertical offset from the Galactic plane is $Z_{\mathrm{Gal},\,\odot} = \qty[]{20.8}{\parsec}$ \citep{2019MNRAS.482.1417B}.

For both the stellar and orbital parameters, we use the 68\% credible interval throughout our analysis, and sometimes approximate the posteriors as Gaussian, taking half this interval as the standard deviation. The uncertainties on $\tau$ and $R_\mathrm{g}$ are shown in Figure \ref{fig:sig_age_Rg}.

\subsection{Thin disc selection}
\label{ssec:DiscSel}

In this work, we restrict our analysis to the thin disc, where the stellar orbits have been less disturbed by dynamical processes (either rapid `non-adiabatic heating' which can be caused by interactions with other galaxies or scattering from giant molecular clouds, or `adiabatic heating' from the secular evolution of the disc \citep[][and references therein]{2019ApJ...878...21T}). The rest of the sample will be studied in future works where we explore other abundance gradients, vertical trends, and stars from the Galactic halo. We select thin disc members from our sample in two stages: first, we make a cut in the [$\alpha$/M] abundance to remove the $\alpha$-enhanced population, which is usually associated with the halo and thick disc (Figure \ref{fig:thin_sel}). We use [$\alpha$/M] as a proxy for [$\alpha$/Fe] as the differences between them\footnote{defining [$\alpha$/Fe] as an average of [X/Fe] for O, Mg, Si, S, and Ca} are negligible. Second, we make a cut to retain stars with $Z_\mathrm{max} < \qty[]{0.5}{\kilo\parsec}$ which are on kinematically cool orbits. This cut is chosen to be slightly wider than the scale height of the thin disc, to accommodate the flaring of the disc in the outer galaxy \citep{2006A&A...451..515M}. Though this cut changes the metallicity distribution of our sample, our results are not very sensitive to its exact position (see Appendix \ref{app:zmax} for details).

\begin{figure}
	\includegraphics[width=\linewidth]{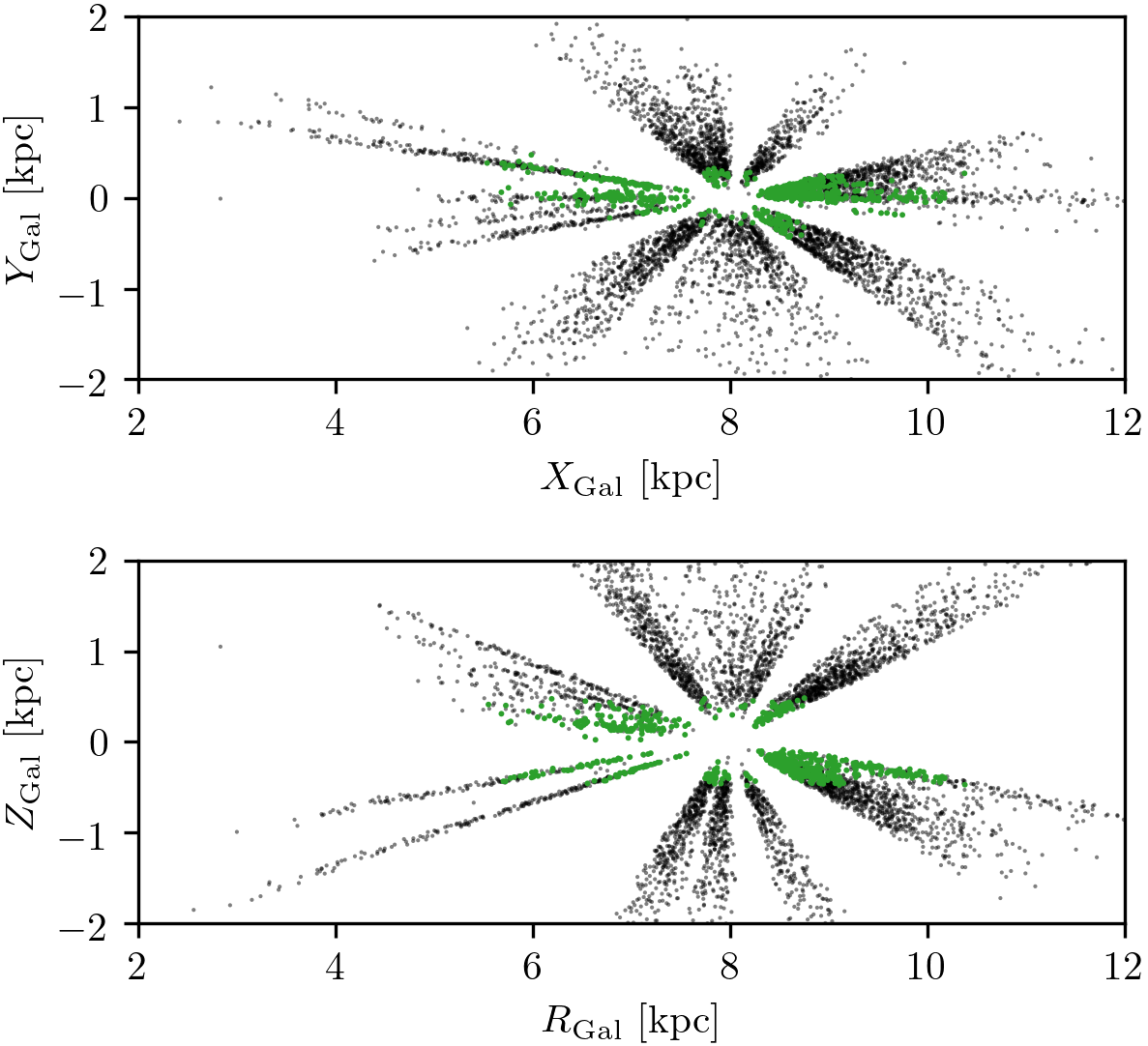}
    \caption{Location of stars in our sample in left-handed Galactocentric cartesian coordinates (\textit{top}) and Galactocentric cylindrical coordinates (\textit{bottom}). The green points show the stars in the thin disc, and the full sample is shown in the background (black points).}
    \label{fig:XY_RZ}
\end{figure}

\begin{figure}
	\includegraphics[width=\linewidth]{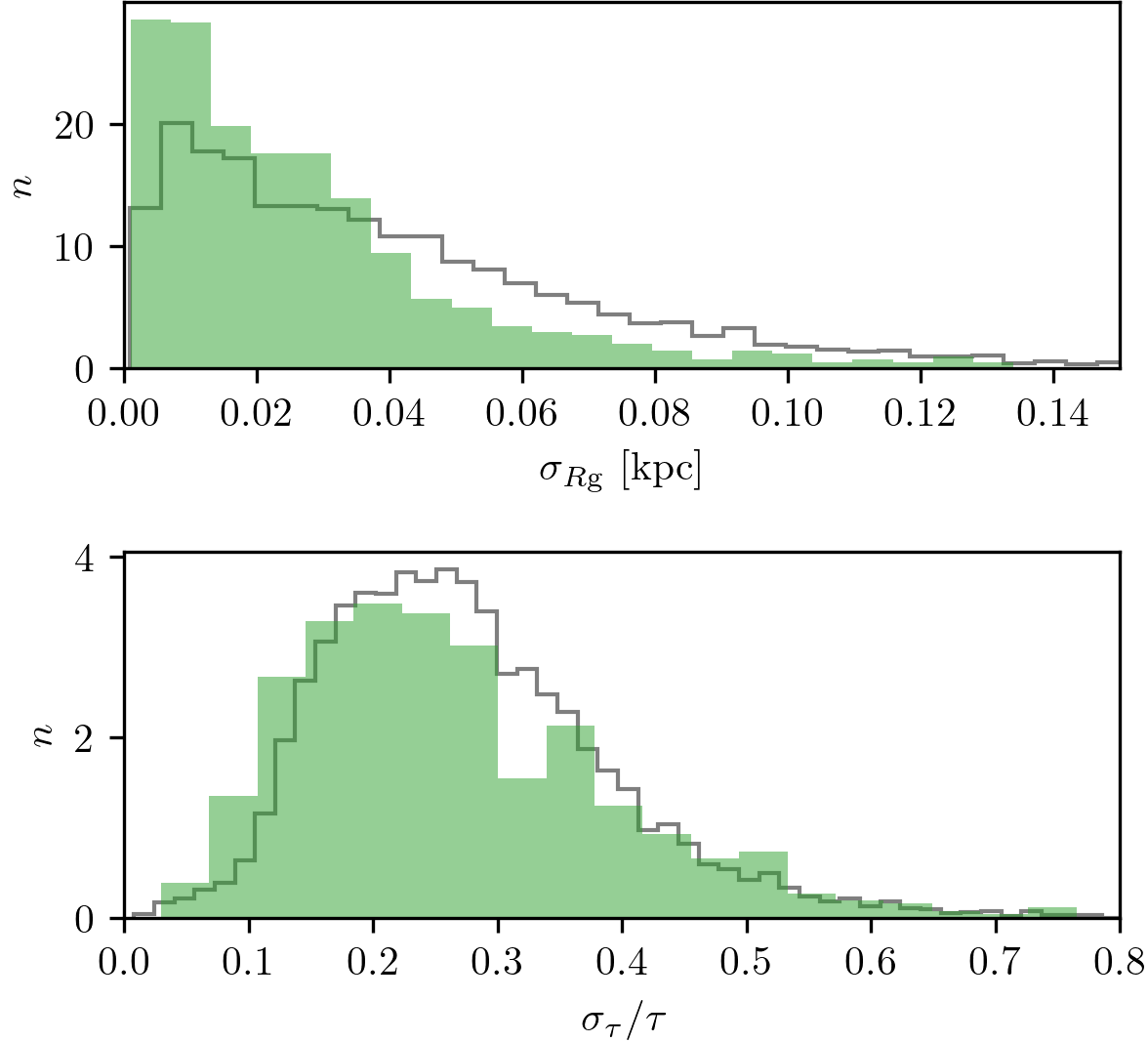}
    \caption{Distribution of the uncertainty on guiding radius (\textit{top}) and fractional uncertainty on age (\textit{bottom}) in the full sample (grey line) and thin disc (filled green). The area under each histogram integrates to one.}
    \label{fig:sig_age_Rg}
\end{figure}

\begin{figure}
	\includegraphics[width=\linewidth]{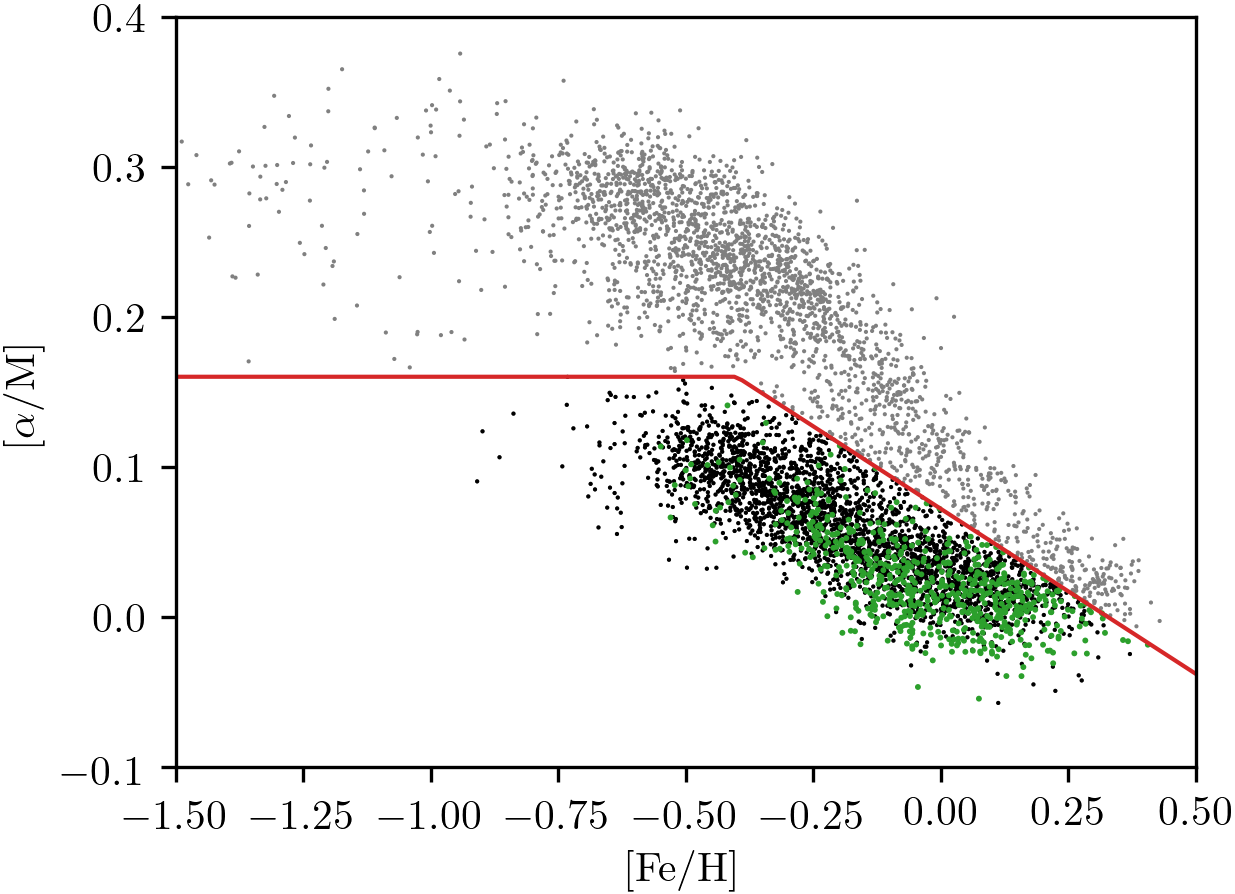}
    \caption{Average $\alpha$ element abundance [$\alpha$/M] vs. [Fe/H]. The high- and low-$\alpha$ sequences were separated along the red line, with the high-$\alpha$ population (grey) removed from the analysis. of the remaining low-$\alpha$ population, stars with $Z_\mathrm{max} \geq \qty[]{0.5}{\kilo\parsec}$ were also removed (black). The final thin disc sample is shown in green.}
    \label{fig:thin_sel}
\end{figure}

Our thin disc sample contains a total of 668 stars and is shown in green in Figures \ref{fig:XY_RZ}, \ref{fig:sig_age_Rg} and \ref{fig:thin_sel}, with median uncertainties $\sigma_\mathrm{[Fe/H]} = 0.007$, $\sigma_\tau / \tau = 0.24$ and $\sigma_{R_\mathrm{g}} = \qty[]{0.02}{\kilo\parsec}$ - though we remind the reader of the comment on $\sigma_\mathrm{[Fe/H]}$ in Section \ref{ssec:Sample_obs}.

\section{The metallicity gradient in bins of age}
\label{sec:Bins}

We investigate the metallicity as a function of $R_\mathrm{g}$ across six bins in age: 
\begin{alignat*}{3}
                        \tau& < \qty[]{1}{Gyr},  &&      ~~\qty[]{1}{Gyr}  \leq \tau&& < \qty[]{2}{Gyr}, \\
    \qty[]{2}{Gyr} \leq \tau& < \qty[]{4}{Gyr},  &&      ~~\qty[]{4}{Gyr}  \leq \tau&& < \qty[]{6}{Gyr}, \\
    \qty[]{6}{Gyr} \leq \tau& < \qty[]{10}{Gyr}, &\quad&   \qty[]{10}{Gyr} \leq \tau&&
\end{alignat*}
where we use bins of increasing width to accommodate the wider posteriors at older ages. For ease of comparison, these bins are the same as those used by \citet{2017A&A...600A..70A}.

We utilise a Bayesian fitting procedure to constrain the metallicity gradient, using a `linear + Gaussian spread' model. In physical terms, this means that we expect stars to be normally distributed about some line intrinsically, rather than having been drawn from a narrow relationship and subsequently scattered, so 
\begin{equation}
    \mathrm{[Fe/H]} \sim \mathcal{N}(m R_\mathrm{g} + c, V).
\end{equation}
This model has three parameters $\phi = {m, c, V}$, where $m$ and $c$ are the linear gradient and offset from $\mathrm{[Fe/H]} = 0$ at $R_\mathrm{g} = 0$, and $V$ is the variance of the Gaussian, oriented perpendicular to the direction of the linear component. Using Bayes' theorem, the posterior probability density function (PDF) of these model parameters given some data $d$ is
\begin{equation}
    p(\phi|d) \propto p(\phi) p(d|\phi),
\end{equation}
where $p(\phi)$ is the prior PDF of the model parameters and $p(d|\phi) = \mathcal{L}$ is the likelihood of observing the data, given the model.

Our fitting procedure follows \citet{2010arXiv1008.4686H} and accounts for uncertainty on both $R_\mathrm{g}$ and [Fe/H] by utilising the full covariance matrix of each star $i$, $\varsigma_i$, in the likelihood. In addition, we reparametrise the problem so that we do not directly constrain $m$ and $c$, but rather the angle to the horizontal, $\cos(\theta) = {1}/{\sqrt{1+m^2}}$, and $c_\perp = c \cos(\theta)$, which avoids our wide priors placing too much emphasis on gradients close to vertical. This gives a new set of parameters $\varphi = \left[\theta, c_\perp, V\right]$. Projecting $\varsigma_i$ to an orthogonal variance $\Sigma_i^2$, we may write the log-likelihood as:
\begin{equation}
    \ln\left(\mathcal{L}\right) \propto - \frac{1}{2} \sum_i \left[  \ln\left(\Sigma_i^2 + V\right) + \frac{\Delta_i^2}{\Sigma_i^2 + V} \right],
\end{equation}
where $\Delta_i$ is the orthogonal displacement of star $i$ from the line. We choose very wide, uniform priors on the parameters $\varphi$, requiring:
\begin{align*}
     -\pi & < \theta < \pi, \\
    -10^3 & < c_\perp < 10^3, \\
        0 & \leq V < 10^4.
\end{align*}

\begin{figure*}
	\includegraphics[width=0.9\linewidth]{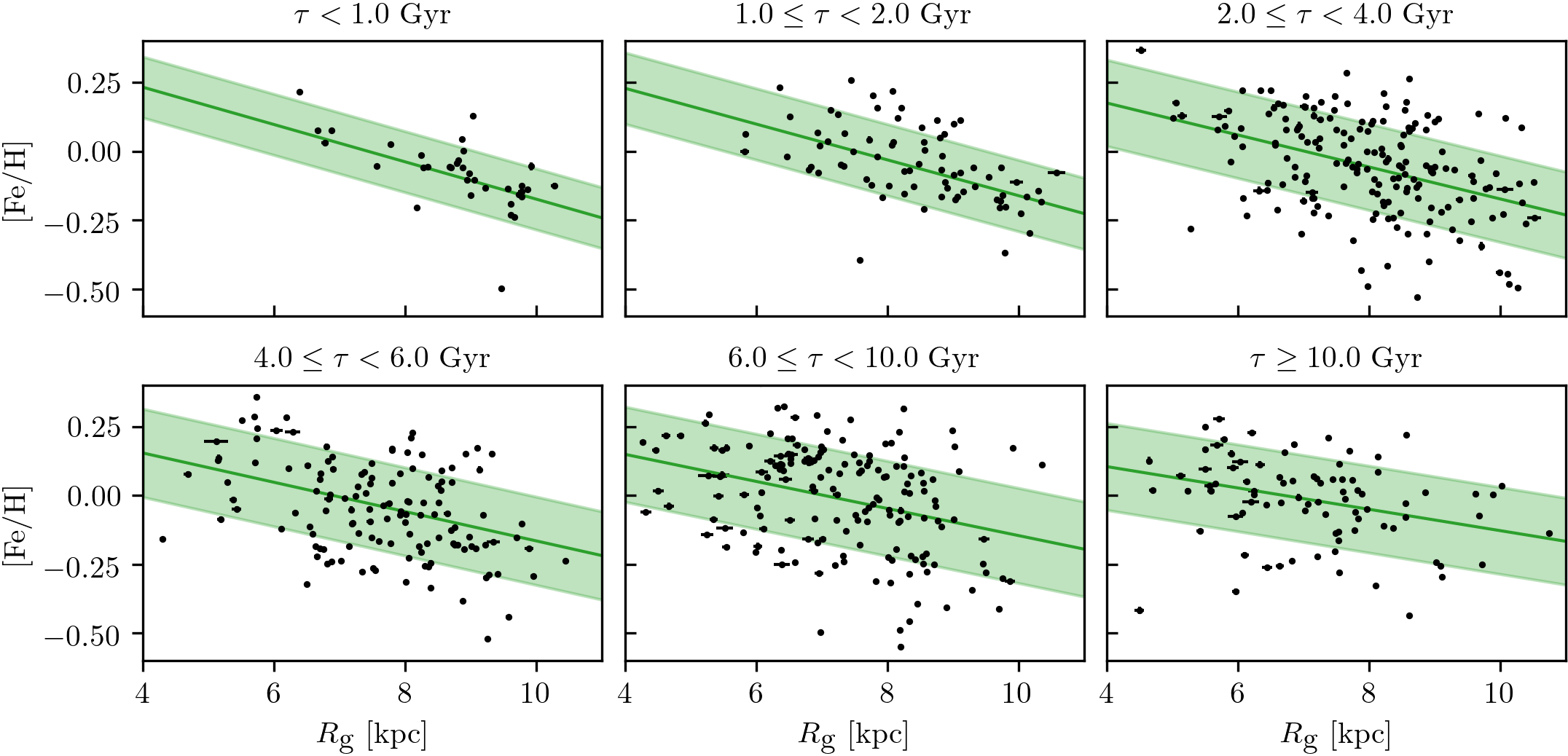}
    \caption{[Fe/H] vs. $R_\mathrm{g}$ in six bins of age. The thin disc sample is shown in black points, binned according to the median age of each star. The green line shows the best-fitting radial metallicity gradient from the MCMC analysis, with a filled region representing $1\sigma$ of the intrinsic spread.}
    \label{fig:bins}
\end{figure*}

\begin{figure}
	\includegraphics[width=\linewidth]{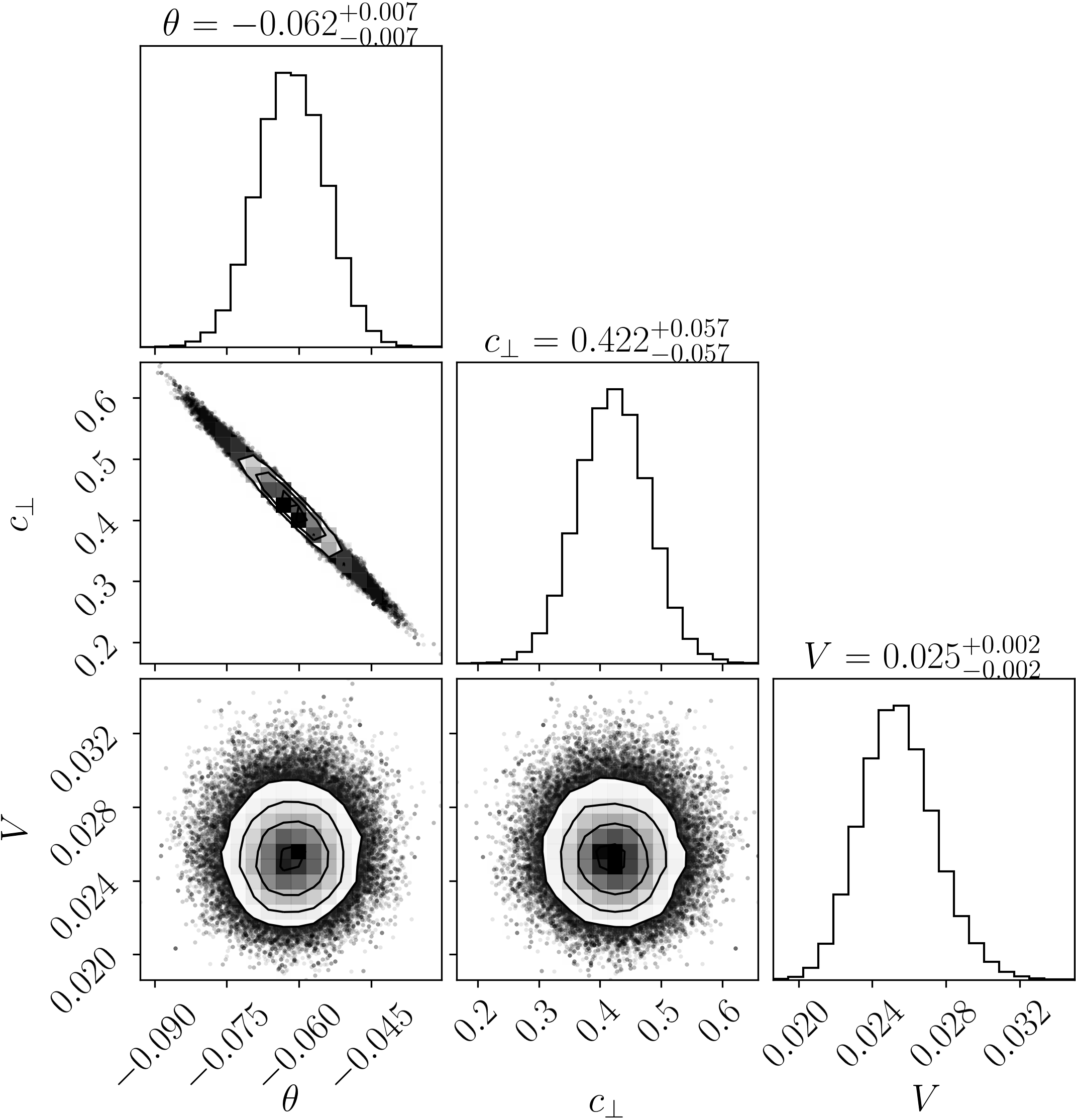}
    \caption{Corner plot \citep{2016JOSS....1...24F} showing the joint and marginalised sampled posterior distributions of the fit parameters $\varphi$ in the age bin $\qty[]{2}{Gyr} \leq \tau < \qty[]{4}{Gyr}$ for one step in the MC (one realisation of $\tau$).}
    \label{fig:corner}
\end{figure}

We explore this parameter space by sampling according to the Markov Chain Monte Carlo (MCMC) algorithm, implemented in \texttt{emcee} \citep{2013PASP..125..306F}. We use 32 walkers which each make 5000 steps, with a burn-in time which is dynamically determined to be 10 times the auto-correlation time (removing around 500 samples per walker). To account for the uncertainty on $\tau$, we perform an overarching Monte Carlo (MC) analysis by repeating this procedure for 500 realisations of $\tau$ which are generated by approximating the posterior distribution from PARAM as Gaussian. This allows the stars to move between bins in each step of the MC. We combine all the post-burn-in samples and report the median and 68\% credible interval as the best-fitting values of the parameters. The age uncertainty contributes approximately 30\% to the overall error budget on these results. The lines corresponding to these values are shown in Figure \ref{fig:bins}, where the filled region indicates the intrinsic spread, $\sqrt{V}$, and the points appear in the bin to which the median age of the star belongs. An example corner plot showing the joint and marginal distributions of the fit parameters $\varphi$ in one bin, for one realisation of $\tau$ is shown in Figure \ref{fig:corner}.

We report the best fitting values of the gradient, $m$, metallicity offset at \qty[]{8}{\kilo\parsec}, $b = m\times8 + c$, and [Fe/H] component of the spread $S = \sqrt{V}\cos(\theta)$ in the left-hand side of Table \ref{tab:res}. The gradient flattens monotonically with age, and the spread is smallest for the youngest populations, increasing until $\tau \approx \qty[]{10}{\giga\year}$. The metallicity at \qty[]{8}{\kilo\parsec} is slightly sub-solar at all ages.

\begin{table*}
    \caption{The median and 68\% credible interval for each parameter from the MCMC analysis in bins of age (\textit{left}) and the HBM, evaluated at the median age, $\tau_\mathrm{med}$ of each bin population (\textit{right}); $m$ is the gradient, $b$ is the [Fe/H] offset at $R_\mathrm{g} = \qty[]{8}{\kilo\parsec}$, and $S$ is the component of the intrinsic spread in the [Fe/H] direction.}
	\label{tab:res}
    \begin{tabular}{rlr|lll||r|lll}
    \toprule
    \multicolumn{1}{l}{} &
       &
      \multicolumn{1}{l|}{} &
      \multicolumn{3}{c||}{MCMC Parameters} &
      \multicolumn{1}{l|}{} &
      \multicolumn{3}{c}{HBM Parameters} \\
    \multicolumn{2}{l}{bin range {[}Gyr{]}} &
      $N_\mathrm{stars}$ &
      $m$ {[}dex/kpc{]} &
      $b$ {[}dex{]} &
      $S$ {[}dex{]} &
      $\tau_\mathrm{med}$ {[}Gyr{]} &
      $m$ {[}dex/kpc{]} &
      $b$ {[}dex{]} &
      $S$ {[}dex{]} \\ 
      \midrule
    \multicolumn{2}{r}{$\tau < 1~~$} &
      35 &
      $-0.068 \substack {+0.017 \\ -0.016}$ &
      $-0.039 \substack {+0.022 \\ -0.023}$ &
      $0.111 \substack {+0.015 \\ -0.013}$ &
      0.7 &
      $-0.078 \substack {+0.011 \\ -0.011}$ &
      $-0.022 \substack {+0.016 \\ -0.015}$ &
      $0.115 \substack {+0.010 \\ -0.009}$ \\
    \multicolumn{2}{r}{$1 \leq \tau < 2~~$} &
      74 &
      $-0.065 \substack {+0.013 \\ -0.013}$ &
      $-0.033 \substack {+0.015 \\ -0.015}$ &
      $0.130 \substack {+0.012 \\ -0.011}$ &
      1.6 &
      $-0.068 \substack {+0.008 \\ -0.007}$ &
      $-0.034 \substack {+0.010 \\ -0.009}$ &
      $0.130 \substack {+0.007 \\ -0.006}$ \\
    \multicolumn{2}{r}{$2 \leq \tau < 4~~$} &
      183 &
      $-0.058 \substack {+0.010 \\ -0.010}$ &
      $-0.058 \substack {+0.011 \\ -0.011}$ &
      $0.157 \substack {+0.008 \\ -0.008}$ &
      3.0 &
      $-0.060 \substack {+0.005 \\ -0.005}$ &
      $-0.044 \substack {+0.006 \\ -0.007}$ &
      $0.145 \substack {+0.005 \\ -0.005}$ \\
    \multicolumn{2}{r}{$4 \leq \tau < 6~~$} &
      132 &
      $-0.053 \substack {+0.012 \\ -0.012}$ &
      $-0.060 \substack {+0.016 \\ -0.016}$ &
      $0.160 \substack {+0.011 \\ -0.010}$ &
      4.9 &
      $-0.054 \substack {+0.005 \\ -0.005}$ &
      $-0.052 \substack {+0.007 \\ -0.007}$ &
      $0.157 \substack {+0.004 \\ -0.004}$ \\
    \multicolumn{2}{r}{$6 \leq \tau < 10$} &
      165 &
      $-0.049 \substack {+0.012 \\ -0.011}$ &
      $-0.048 \substack {+0.016 \\ -0.016}$ &
      $0.173 \substack {+0.011 \\ -0.0010}$ &
      7.6 &
      $-0.048 \substack {+0.006 \\ -0.006}$ &
      $-0.059 \substack {+0.009 \\ -0.009}$ &
      $0.169 \substack {+0.006 \\ -0.006}$ \\
    \multicolumn{2}{r}{$10 \leq \tau~~~~~~~~~$} &
      87 &
      $-0.039 \substack {+0.013 \\ -0.013}$ &
      $-0.051 \substack {+0.020 \\ -0.020}$ &
      $0.158 \substack {+0.012 \\ -0.012}$ &
      12.6 &
      $-0.041 \substack {+0.008 \\ -0.008}$ &
      $-0.067 \substack {+0.012 \\ -0.013}$ &
      $0.184 \substack {+0.010 \\ -0.009}$ \\ 
      \bottomrule
    \end{tabular}
\end{table*}

\section{A hierarchical Bayesian approach}
\label{sec:HBM}

Reducing a dataset to a small number of bins can lead to misleading and sometimes unreliable results. In the specific context of this work, the intrinsic spread around the linear relationship carries essential information, so conflating it with a scatter caused by the mix of ages in a bin makes robust inference a challenge. This is compounded by the fact that the bins are wider, and therefore the populations are a worse approximation of mono-age, at older ages. In this section, we show that, with our large dataset and some simplifying assumptions, it is possible to remove the need for binning altogether.

To constrain the changes in the metallicity distribution with age we constructed a hierarchical Bayesian model (HBM). This allows the fit parameters to vary for each star (they are now the latent parameters), constrained by some population-level rule which describes how they vary with age. We reparameterise to describe the variation with $\log(\tau)$ instead because the increasing age uncertainty as a function of age makes the parameter space difficult to explore. In this case, we assume a linear relationship for $m$ and $b$ with $\log(\tau)$, and (since the spread must have a strictly non-negative value) $S$ follows an exponential in $\log(\tau)$:
\begin{equation}
    \mathrm{[Fe/H]} \sim \mathcal{N}(m R_\mathrm{g} + \left(b-8m\right), S^2);
\end{equation}
\begin{align*}
    & m = k_{1} A_m \log(\tau) + k_{2} B_m, \\
    & b = k_{3} A_b \log(\tau) + k_{4} B_b, \\
    & S = \exp(k_{5} A_S \log(\tau) + k_{6} B_S).
\end{align*}
The parameters $\Psi = {A_m, B_m, A_b, B_b, A_S, B_S}$ are the population-level hyperparameters and the constants $k_i$ are introduced to rescale each to be close to unity, in order to increase the efficiency of our sampling. These simple relationships are a pragmatic choice, providing a test of the results from the binned analysis for a small number of hyperparameters. 

\begin{figure}
	\includegraphics[width=0.8\linewidth, center]{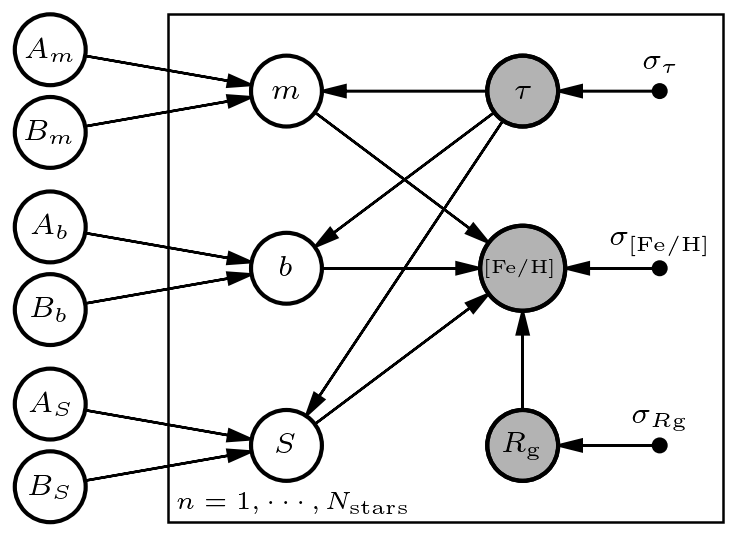}
    \caption{A probabilistic graphical model of the hierarchical Bayesian model (HBM). The properties of individual stars are shown inside the rectangle, and the hyperparameters are shown outside. Grey nodes represent observables and their respective observational uncertainties are given by the small black nodes.}
    \label{fig:PGM}
\end{figure}

Figure \ref{fig:PGM} shows the probabilistic graphical model corresponding to this analysis, where the dependencies between the hyperparameters, latent parameters and observables are indicated by arrows. Bayes' theorem now becomes:
\begin{equation}
    p\left(\Psi, \phi|d\right) \propto p\left(\Psi\right) p\left(\phi|\Psi\right) p\left(d|\phi\right).
\end{equation}
We choose minimally informative priors on the hyperparameters, $\Psi$, with each drawn from a normal of the form $\mathcal{N}(\mu, \sigma^2)$ such that:
\begin{alignat*}{2}
    &A_m \sim \mathcal{N}(3, 9),  &\quad& B_m \sim \mathcal{N}(0, 9), \\
    &A_b \sim \mathcal{N}(-3, 9), &&      B_b \sim \mathcal{N}(0, 9), \\
    &A_S \sim \mathcal{N}(0, 25),  &&      B_S \sim \mathcal{N}(0, 25), \\
\end{alignat*}
and, because the HBM is a generative model, we also require priors on $R_\mathrm{g}$ and $\log\left(\tau\right)$. We choose these to be broad and likelihood-dominated:
\begin{alignat*}{1}
     R_\mathrm{g} &\sim \mathcal{N}(8, 25) \\
     \log\left(\tau\right) &\sim \mathcal{N}(0.7, 0.09).
\end{alignat*}

We explore this new parameter space with the No-U-Turn-Sampler (NUTS) variant of the MCMC algorithm \citep{2011arXiv1111.4246H}, implemented in \texttt{NumPyro} \citep{bingham2019pyro, 2019arXiv191211554P}, using five chains which each make a total of 15000 samples, 5000 of which are reserved for warm-up. We confirm model convergence by ensuring $\hat{r} < 1.04$ for all parameters \citep{1992StaSc...7..457G}.

\begin{figure}
	\includegraphics[width=\linewidth]{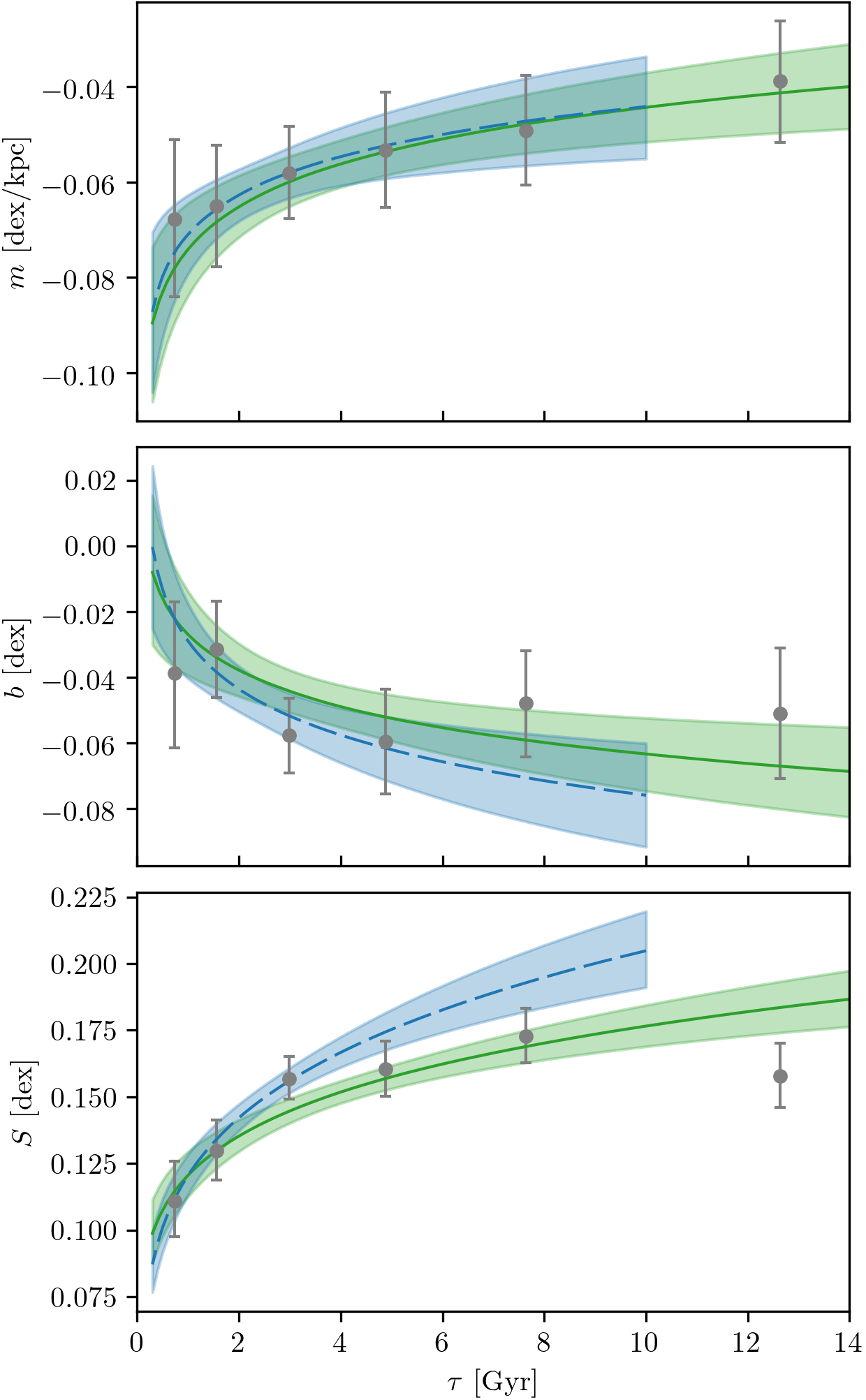}
    \caption{Evolution of the fit parameters as a function of stellar age. The grey points show the best-fitting value and 68\% credible interval in each bin from the MCMC analysis (plotted against the median age of the population in the bin). The green line and filled region show the result and 68\% credible interval from the HBM. The dashed blue line and filled region show the HBM result when stars older than \qty[]{10}{\giga\year} are excluded from the analysis.}
    \label{fig:HBM}
\end{figure}

The results for this analysis are shown in Figure \ref{fig:HBM}, where the line indicates the median value of the posterior distribution and the filled band is the 68\% credible interval, which is larger for younger and older stars where the data are sparser. For comparison, we show the results of the age-binned analysis, plotted against the median age of the population in each bin, with error bars which show the 68\% credible interval. The results of the HBM at the median age of each bin population are reported in Table \ref{tab:res}, and both sets of results are discussed more fully in the next section.

\section{Discussion}
\label{sec:Discussion}

We have shown results from two complementary methods to measure the radial metallicity gradient over the history of the MW's disc. In the next section, we discuss these results in more detail, as well as what can be learned from their comparison. We give special consideration to the oldest stars in Section \ref{ssec:old_stars} and comment on the possible biases present in our work in Section \ref{ssec:sel_eff}. Finally, in Section \ref{ssec:comp_lit} we compare to predictions from theoretical models of chemo-dynamical evolution and set our work in the context of the literature.

\subsection{Results of this work}
\label{ssec:these_results}

Our sample of K2 oscillating red giants provides excellent coverage of a wide range of guiding radii and ages, which allows the metallicity evolution of the thin disc to be investigated. Using hierarchical Bayesian analysis we obtain constraints which are free from the additional uncertainties introduced by binning. However, in this analysis, it is necessary to assume a functional form for the evolution of the fit parameters, and therefore the HBM does not have the freedom to trace all features of the data. In this way the MCMC analysis performed on the binned data and the HBM are complementary and it is useful to compare their results in detail.

As shown in Figure \ref{fig:HBM} and Table \ref{tab:res}, the two methods agree very well (within $1 \sigma$) at all ages for the gradient and offset at \qty[]{8}{\kilo\parsec} of the radial metallicity relationship. The largest differences are in the stars younger than \qty[]{2}{\giga\year}, which may be due to the smaller sample size or the assumed functional forms within the HBM. We find that $b$ is below solar metallicity at all ages, suggesting that the Sun was born closer to the Galactic centre than its current position, in agreement with \citet{1996A&A...314..438W, 2013A&A...558A...9M, 2018MNRAS.481.1645M} and \citet{2018ApJ...865...96F}.

There are more significant differences between the results for the intrinsic spread. This is not surprising because, in the MCMC analysis, the spread arising from the width of each bin and the sample size within each bin are conflated with the underlying physics. Overall the agreement between the methods suggests that the effects of the binning do not dominate. This verification is important as understanding this spread is an important step towards a measurement of the effects of churning \citep[e.g.][]{2008MNRAS.388.1175H, 2015A&A...580A.127K}{}{}, though quantifying the effects of radial migration is beyond the scope of this study. It seems likely that the difference in population size (Table \ref{tab:res}) is the cause of the small fluctuations around the trend shown by the HBM, with slightly larger populations correlating with a larger spread. The deviation is largest in the population older than \qty[]{10}{\giga\year} and, while the smaller sample size in this bin is still relevant, these stars merit special consideration because they were likely born in an epoch prior to the formation of the thin disc.

\subsection{Stars older than \qty[]{10}{\giga\year}}
\label{ssec:old_stars}

Models of the assembly history of the MW show that the Galactic disc formed $\approx \qty[]{10}{\giga\year}$ ago \citep[e.g.][]{2022MNRAS.514..689B}{}{}. Therefore stars older than this were not born in the thin disc as we know it today, and this bin contains the highest fraction of contaminants from the debris of mergers or other processes in the MW before the onset of disc formation. Since it is necessary to assume a functional form for the evolution of $S$, the HBM is not able to trace all features present in the data, so the difference between it and the binned result does not necessarily indicate an effect introduced by the binning. However, we can test the effect of the old population in the HBM by removing it and rerunning the analysis. 

After removing stars with $\tau > \qty[]{10}{\giga\year}$, the HBM results for the slope and offset are consistent within their uncertainties (the blue lines in Figure \ref{fig:HBM}), but we find a significantly steeper increase in the spread with age. This indicates that the reduced spread is, at least partly, a physical effect because the old population flattens the fit of the HBM overall. This could be a signature of the small radial extent and centrally concentrated star formation in that epoch of the MW. We also test whether the difference in these HBM results could be due to the the larger uncertainty on the age of the older stars, but find that the effect is independent of the size of the age uncertainty.

There is a disagreement between these results and the intermediate age bins ($\qty[]{4}{Gyr} \leq \tau < \qty[]{6}{Gyr}$ and $\qty[]{6}{Gyr} \leq \tau < \qty[]{10}{Gyr}$), with the HBM predicting a larger spread, which is driven by the assumption in the HBM of linear dependence on $\log(\tau)$ and the steep increase in spread at young ages.

\begin{figure}
	\includegraphics[width=\linewidth]{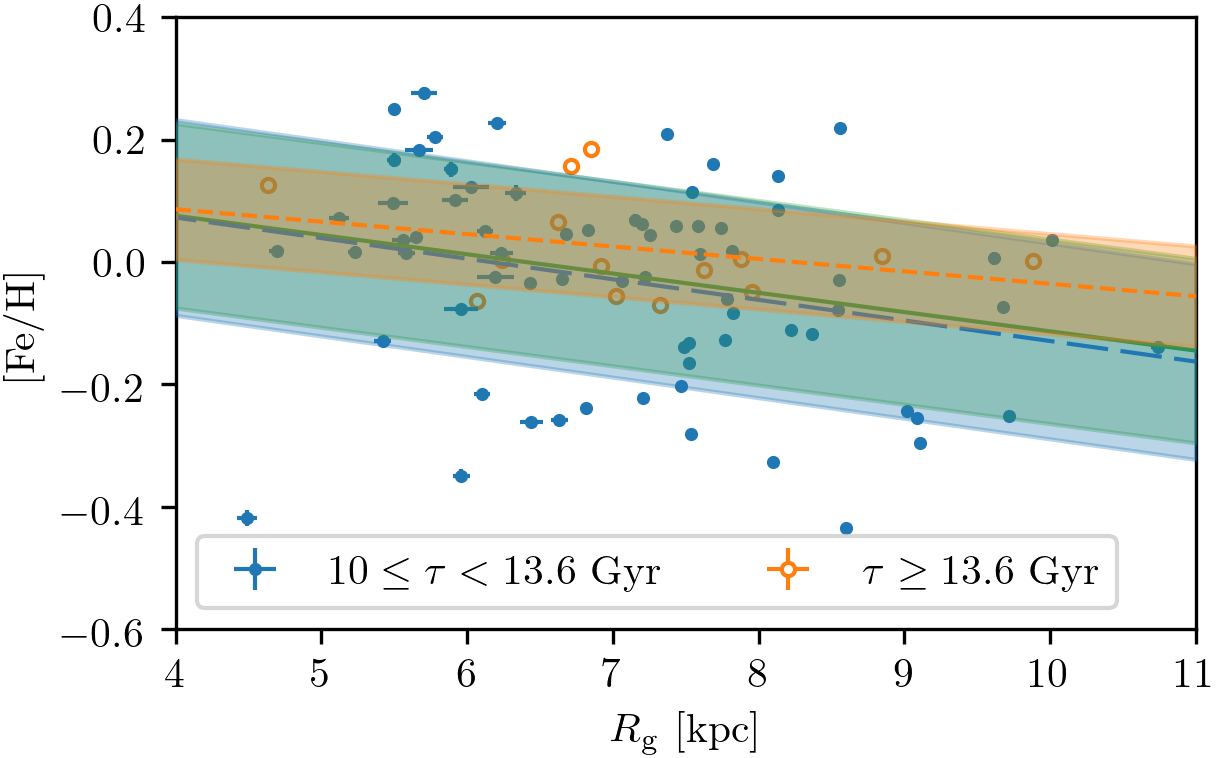}
    \caption{Results of the MCMC analysis for the oldest stars in our sample. The green line and filled region in the background show the result when including all stars with $\tau \geq \qty[]{10}{\giga\year}$, the long-dashed blue line and filled region (which are almost identical to the green) show the result of the analysis excluding stars with $\tau \geq \qty[]{13.6}{\giga\year}$, and the short-dashed orange line and filled region show the result based on only those stars.}
    \label{fig:MCMC_old}
\end{figure}

The age prior in PARAM is deliberately wide and uninformative, which means our sample contains some stars with median age greater than \qty[]{13.6}{\giga\year}. Since the age posteriors are wide, we do not remove these stars from our analysis, but we confirm that they are not dominating the result in the oldest bin by analysing them separately (Figure \ref{fig:MCMC_old}). We find that, though these stars do have a gradient which is significantly flatter than those with $\qty[]{10}{\giga\year} \leq \tau < \qty[]{13.6}{\giga\year}$, the final fit in the bin is almost unaffected by their presence, being fully consistent to the result with the very old stars removed.

\subsection{Possible sources of bias}
\label{ssec:sel_eff}

In this work, we choose the guiding radius as our primary radial coordinate because it provides a robust estimate of the Galactocentric distance of the star over its whole orbit. For the same sample, this will naturally tend to flatten the radial metallicity gradient because we observe a limited range of $R_\mathrm{Gal}$. In the inner Galaxy, it's only possible to observe stars with $R_\mathrm{g} < R_\mathrm{Gal,\,min}$ if they are eccentric enough that their apocentre radius $R_\mathrm{ap} \geq R_\mathrm{Gal,\,min}$. We cannot observe stars on circular orbits with the same $R_\mathrm{g}$ because they never enter our observable region, so we spread our sample over a wider range of radii, without increasing the range in metallicity.  We show this in Appendix \ref{app:RGal}, Figures \ref{fig:bins_Rgal} and \ref{fig:HBM_binsRgal}, were for $\tau > \qty[]{2}{\giga\year}$ we do find a steeper gradient with $R_\mathrm{Gal}$. For $\tau < \qty[]{2}{\giga\year}$ the effect is reversed, which is likely to be because the flattening effect of $R_\mathrm{g}$ is smaller due to the lower eccentricities of the younger population. The irregular distribution of the stars in $R_\mathrm{Gal}$ (Figure \ref{fig:bins_Rgal}) may then lead to a slight flattening of the gradient.

The equivalent argument can be made for stars with $R_\mathrm{g} > R_\mathrm{Gal,\,max}$ and pericentre radius $R_\mathrm{pe} \leq R_\mathrm{Gal,\,max}$, but in our sample the two effects do not appear symmetrically: we have more stars with $R_\mathrm{g} < R_\mathrm{Gal, min}$ than $R_\mathrm{g} > R_\mathrm{Gal,\,max}$. The stellar density in the outer disc is lower, so we would see fewer of such stars in a spatially uniform sample, but we see roughly the same number of stars close to our radial limits ($R_\mathrm{Gal,\,min} \approx \qty[]{5.5}{\kilo\parsec}, R_\mathrm{Gal,\,max} \approx \qty[]{10}{\kilo\parsec}$). We can motivate the asymmetry by considering the probability of observing the `inner' star ($R_\mathrm{g} < R_\mathrm{Gal,\,min}$) while it is passing through the observable region, compared to the `outer' star. As an example, our sample contains one star with $R_\mathrm{g} < \qty[]{4}{\kilo\parsec}$ - i.e. which is observed more than \qty[]{1.5}{\kilo\parsec} away from its guiding radius. This means it has an eccentricity of $e \gtrsim 0.3$. An equivalent target in the outer disc with $R_\mathrm{g} > \qty[]{11.5}{\kilo\parsec}$ has $e \gtrsim 0.2$. Since low eccentricity stars dominate the population in the thin disc, it might seem surprising that we do not observe any stars with $R_\mathrm{g} \gtrsim \qty[]{10.5}{\kilo\parsec}$. However, when we consider the probability of each star being in the part of its orbit where it is close enough to the solar position to be observable, as a fraction of its full orbit, we find that $P_\mathrm{obs,\,inner} \approx 2P_\mathrm{obs,\,outer}$. Given the small numbers involved, this is enough to explain the asymmetry in our sample.

Working in $R_\mathrm{g}$ may also introduce a kinematic bias since higher eccentricity stars tend to be older and thus more metal-poor. This could mean that we miss higher metallicity stars at low $R_\mathrm{g}$ and recover an overly flat gradient. Examining Figure \ref{fig:bins}, this does not appear to be the case, but as a test we repeat the HBM analysis, removing targets with $R_\mathrm{g} < \qty[]{5.5}{\kilo\parsec}$. The result is shown in Figure \ref{fig:HBM_Rg55}, where we see a slightly steeper gradient for stars older than \qty[]{2}{\giga\year}, but it is still fully consistent with the results from the full sample within the uncertainties.

\subsection{Comparison to literature}
\label{ssec:comp_lit}

In this section, we compare the results of our analysis to those from other studies based on field stars, Cepheid variables and OCs, as well as theoretical models.

\subsubsection{Field stars and Cepheid Variables}
Below we compare only to works with ages which are `directly' inferred, either from comparison to isochrones or asteroseismology, similar to those used in this work. A comparison to the numerous age estimates available from machine learning methods is beyond the scope of this work, and will be completed in a later study.

\citet{2004A&A...418..989N} and \citet{2011A&A...530A.138C} investigated the radial metallicity gradient in the Geneva-Copenhagen Survey (GCS). The large sample of local solar neighbourhood F and G dwarfs extends to $\approx \qty[]{0.4}{\kilo\parsec}$ from the solar position and has metallicity determinations based on Str\"{o}mgren photometry and ages obtained from comparison to isochrones. Though the precision of the ages is around 25\%, they are subject to significant uncertainties arising from the calibration of the metallicity and temperature scales. The sample is dominated by young stars ($\tau \approx \qty[]{2}{\giga\year}$) due to the magnitude limits of the survey and its local volume, and between the works of \citeauthor{2004A&A...418..989N} and \citeauthor{2011A&A...530A.138C} the ages of many stars change by $\approx \qty[]{1}{\giga\year}$ as a result of changes to these calibrations. 

\citet{2004A&A...418..989N} use the mean radius $R_\mathrm{m} = 0.5 \times (R_\mathrm{ap} + R_\mathrm{pe})$, extending their radial range to $\qty[]{6}{\kilo\parsec} < R_\mathrm{m} < \qty[]{9}{\kilo\parsec}$, and find a gradient of $-0.076 \pm \qty[per-mode=symbol]{0.014}{\dex\per\kilo\parsec}$ for stars younger than \qty[]{1.5}{\giga\year}, and a steeper gradient of $0.099 \pm \qty[per-mode=symbol]{0.011}{\dex\per\kilo\parsec}$ in an intermediate age population of $\qty[]{4}{\giga\year} < \tau < \qty[]{6}{\giga\year}$. Their oldest population ($\tau > \qty[]{10}{\giga\year}$) is consistent with no gradient ($0.028 \pm \qty[per-mode=symbol]{0.036}{\dex\per\kilo\parsec}$) and they suggest that this indicates that these stars are not thin disc members. We discuss the stars older than \qty[]{10}{\giga\year} present in our sample in Section \ref{ssec:old_stars}.

\citet{2011A&A...530A.138C} improved on the age estimates of \citeauthor{2004A&A...418..989N} and reanalysed the GCS sample to find a metallicity gradient with both $R_\mathrm{m}$ and $R_\mathrm{g}$. They found a similar trend of a steeper gradient at intermediate ages than for the youngest stars, before flattening with age to a positive gradient in the older population. Their gradient with $R_\mathrm{m}$ is \qty[per-mode=symbol]{0.05}{\dex\per\kilo\parsec} -- \qty[per-mode=symbol]{0.01}{\dex\per\kilo\parsec} less (in the sense of flattening a negative gradient) than \citeauthor{2004A&A...418..989N} across the age range, and the main effect of switching to $R_\mathrm{g}$ is to soften the trend towards a steeper gradient at intermediate ages so that it is nearly flat.

\citet{2017A&A...600A..70A} used a sample of around 400 red giants observed by CoRoT (from which they obtain asteroseismic age constraints) and APOGEE, with $\qty[]{6}{\kilo\parsec} < R_\mathrm{Gal} < \qty[]{13}{\kilo\parsec}$ and $|Z_\mathrm{Gal}| < \qty[]{0.3}{\kilo\parsec}$. After applying a correction for the selection function (derived from comparison to a chemo-dynamical simulation) they find a gradient with $R_\mathrm{Gal}$ of $-0.058 \pm 0.008 \pm \qty[per-mode=symbol]{0.003}{\dex\per\kilo\parsec}$ (statistical and systematic uncertainty) for $\tau < \qty[]{1}{\giga\year}$, steepening to $-0.066 \pm 0.007 \pm \qty[per-mode=symbol]{0.002}{\dex\per\kilo\parsec}$ between \qty[]{1}{\giga\year} and \qty[]{4}{\giga\year}, before flattening to $\approx \qty[per-mode=symbol]{0.03}{\dex\per\kilo\parsec}$ between \qty[]{6}{\giga\year} and \qty[]{10}{\giga\year}.

Finally, \citet{2015RAA....15.1209X, 2019MNRAS.482.2189W} and \citet{2021ApJ...922..189V} used large samples of main sequence turn-off and subgiant stars (0.3 million, with $\qty[]{8}{\kilo\parsec} < R_\mathrm{Gal} < \qty[]{11}{\kilo\parsec}$; 0.94 million, with $\qty[]{7}{\kilo\parsec} < R_\mathrm{Gal} < \qty[]{11}{\kilo\parsec}$; and 1.3 million, with $\qty[]{6}{\kilo\parsec} < R_\mathrm{g} < \qty[]{10}{\kilo\parsec}$, respectively) observed by LAMOST, with ages determined from isochrone comparisons. \citeauthor{2015RAA....15.1209X} and \citeauthor{2019MNRAS.482.2189W} selected stars in slices of $|{Z_\mathrm{Gal}}|$ and investigated the gradient with $R_\mathrm{Gal}$. For stars close to the plane, both find the steepest gradient at intermediate ages (\qty[]{6}{\giga\year} - \qty[]{8}{\giga\year} in \citeauthor{2015RAA....15.1209X} and \qty[]{4}{\giga\year} - \qty[]{6}{\giga\year} in \citeauthor{2019MNRAS.482.2189W}), with younger and older stars showing shallower gradients. The gradients of \citeauthor{2019MNRAS.482.2189W} are consistent with our results for the youngest and oldest stars, and steepen to $\approx \qty[per-mode=symbol]{-0.1}{\dex\per\kilo\parsec}{}$ in between. \citeauthor{2021ApJ...922..189V} select stars with low $Z_\mathrm{max}$ and $\alpha$-abundance (similar to the procedure in this work) and find that the gradient with $R_\mathrm{g}$ flattens monotonically with age. Their results agree very well with this work at all ages.
 
Though most of the results described above agree with each other, and this work, within a few $\sigma$, they fall into two groups according to how the gradient varies with age. In this work and in \citet{2021ApJ...922..189V} the gradient flattens smoothly stellar age, while the other studies show a steeper gradient in intermediate-age stars of $\tau \approx \qty[]{4}{\giga\year}$ than in the younger population (though this is least obvious in the $R_\mathrm{g}$ case of \citet{2011A&A...530A.138C}). comparisons are complicated by the different selection functions of each sample, but below we consider two other possible causes of this difference: the degree to which the population is contaminated by thick disc stars, and the choice of radial coordinate.

Considering first the contamination by thick disc stars, we refer to the works of \citet{2014A&A...564A.115A} (using $R_\mathrm{g}$ and $Z_\mathrm{max}$) and \citet{2014AJ....147..116H} (using $R_\mathrm{Gal}$ and $Z_\mathrm{Gal}$), in addition to \citet{2015RAA....15.1209X, 2019MNRAS.482.2189W} and \citet{2021ApJ...922..189V}, on the vertical dependence of the metallicity gradient. They all found that the radial metallicity gradient flattens with height above the Galactic plane. Including these thicker-disc stars could, therefore, flatten the overall gradient. This effect would be especially important in the younger population because the intrinsic scatter is low and, when working in $R_\mathrm{g}$ or $R_\mathrm{m}$, because the radial extent tends to be reduced due to the generally low eccentricities compared to the older stars. We show the effect of including stars with higher $Z_\mathrm{max}$ in Appendix \ref{app:zmax} where Figure \ref{fig:HBM_Zcuts} shows a flatter gradient in the youngest age bin, more similar to previous studies, and \citeauthor{2021ApJ...922..189V} also recover this effect for intermediate $Z_\mathrm{max}$ ($\qty[]{0.25}{\kilo\parsec} < Z_\mathrm{max} < \qty[]{0.75}{\kilo\parsec}$). Though the contamination of the GCS sample is likely to be small due to its small range in $R_\mathrm{Gal}$ and $Z_\mathrm{Gal}$, a few stars passing through the thin disc could be enough to cause this effect. The range of the \citet{2017A&A...600A..70A} sample is larger and the selection for thin disc members is based on $Z_\mathrm{Gal}$, so stars on kinematically hot orbits which are currently close to the plane of the disc are still included. Since the studies of \citet{2014A&A...564A.115A} and \citet{2014AJ....147..116H} do not include age information, it is difficult to be sure that the flattening of the gradient in their results is driven by the distance from the plane, rather than an effect of the shift to older age distributions at a greater distance from the disc. However, \citet{2015RAA....15.1209X, 2019MNRAS.482.2189W} and \citet{2021ApJ...922..189V} show this effect across a range of ages, and it will be investigated in our sample in a future study.

The choice of radial coordinate may also play a role: as described in Section \ref{ssec:sel_eff}, we would naively expect the change from $R_\mathrm{Gal}$ to $R_\mathrm{g}$ to flatten the observed metallicity gradient. We show the results of our analysis using $R_\mathrm{Gal}$ in Appendix \ref{app:RGal} where, for stars older than \qty[]{3}{\giga\year}, we do indeed see a steeper gradient. However, the gradient in the younger population is steeper in the $R_\mathrm{Gal}$ results, reproducing the effect seen in GCS and the works of \citet{2015RAA....15.1209X, 2017A&A...600A..70A} and \citet{2019MNRAS.482.2189W}. As mentioned above, almost all of these stars have very low eccentricity, so the flattening effect of using $R_\mathrm{g}$ would be reduced, and it is possible that the uneven radial distribution (Figure \ref{fig:bins_Rgal}) in this small sample results in a steeper gradient overall.

Cepheid variables are good tracers of the present-day metallicity gradient as they provide a long radial baseline for targets which are 20 - \qty[]{400}{\mega\year} old. \citet{2014A&A...566A..37G} compiled a sample of 450 Cepheids with $\qty[]{4}{\kilo\parsec} < R_\mathrm{Gal} < \qty[]{19}{\kilo\parsec}$ and found a gradient of $-0.05$ - \qty[per-mode=symbol]{-0.06}{\dex\per\kilo\parsec}, depending on the sample used and the cut in $Z_\mathrm{Gal}$ applied, which is consistent with our MCMC results for the bin containing the youngest stars.

\subsubsection{Open Clusters}
There are many studies in which the radial metallicity gradient is constrained by observations of OCs \citep[e.g.][]{2009A&A...494...95M, 2017MNRAS.470.4363C, 2019MNRAS.490.1821C, 2021MNRAS.503.3279S, 2022Univ....8...87S, 2022arXiv220605534G, 2022AJ....164...85M, 2022MNRAS.509..421N} Here we mention just three recent examples, which benefit from advances in high-resolution spectroscopy, and both astrometry (for improving membership determination) and all-sky spectral analysis from \textit{Gaia}.

\citet{2022AJ....164...85M} use a sample of 85 OCs from the Open Cluster Chemical Abundances and Mapping survey (OCCAM), based on APOGEE DR17 abundances, with a radial range of $\qty[]{6}{\kilo\parsec} < R_\mathrm{Gal} < \qty[]{16}{\kilo\parsec}$. They find a break in the radial metallicity relationship at around \qty[]{12}{\kilo\parsec}, recovering a gradient of $-0.073 \pm \qty[per-mode=symbol]{0.002}{\dex\per\kilo\parsec}$ inside and $-0.032 \pm \qty[per-mode=symbol]{0.002}{\dex\per\kilo\parsec}$ outside this radius.

\citet{2022MNRAS.509..421N} use 136 OCs compiled from various surveys and find that the relationship is well fit by a single line with gradient \qty[per-mode=symbol]{-0.058}{\dex\per\kilo\parsec} over a similar radial range in $R_\mathrm{Gal}$. Their results are supported by photometric observations of a further 10 clusters, and they suggest that the appearance of a break in the relationship as seen in \citet{2022AJ....164...85M} is strongly dependent on the specific objects included in the study, as well as the adopted distance scale.

Both these studies suffer to some extent from a fairly limited sample size, particularly in resolving the question of a break in the relationship. Spectral analysis from \textit{Gaia} may now start to resolve this, for example, \citet{2022arXiv220605534G} have a sample of 503 clusters. However, they restrict their analysis of the metallicity gradient to $R_\mathrm{Gal} < \qty[]{12}{\kilo\parsec}$, so do not probe the region in question. They find a gradient of $-0.063 \pm \qty[per-mode=symbol]{0.008}{\dex\per\kilo\parsec}$ with $R_\mathrm{g}$ in the inner disc.

\begin{figure*}
	\includegraphics[width=0.9\linewidth]{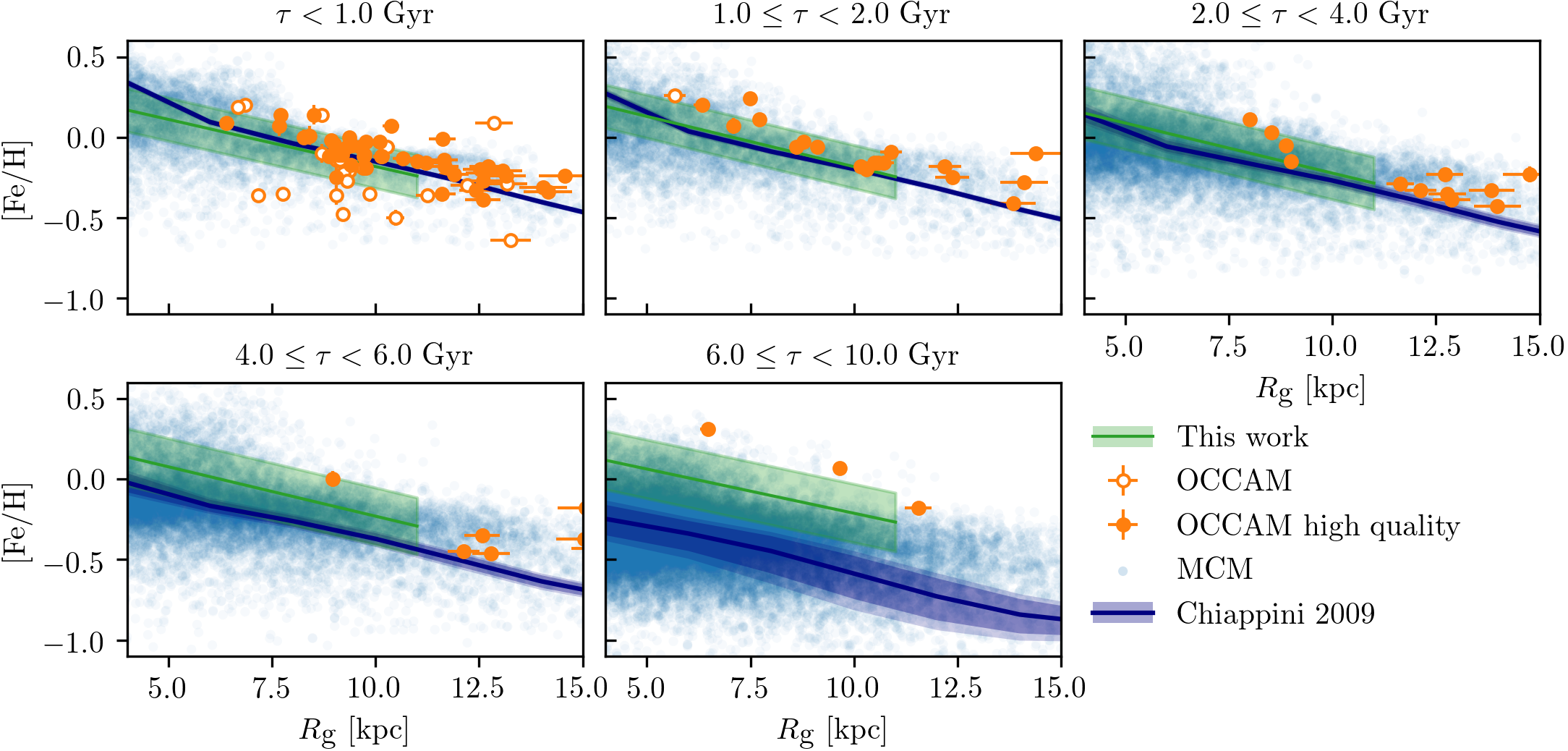}
    \caption{[Fe/H] vs. $R_\mathrm{g}$ in five bins of age. The orange and white points show the high-quality and plausible samples from the OCCAM survey \citep{2022AJ....164...85M}, respectively. The blue points are the results of the MCM chemo-dynamical simulation \citep{2013A&A...558A...9M, 2014A&A...572A..92M}, with the underlying chemical evolution model of \citet{2009IAUS..254..191C} shown by the blue line and filled region. The green line shows the best-fitting radial metallicity gradient from the MCMC analysis of this work, with a filled region representing $1\sigma$ of the intrinsic spread.}
    \label{fig:MCM}
\end{figure*}

Investigating the evolution of the radial metallicity relationship with OCs presents an additional challenge, as the small samples must be broken into bins with few data in each, and the age range is restricted to a few Gyr at most. \citet{2022arXiv220605534G} and \citet{2022MNRAS.509..421N} divide their samples into 4 and 5 age bins, extending up to $\approx 5$ and $\approx \qty[]{7}{\giga\year}$ respectively. In both cases, the gradient is found to increase as a function of age: the opposite of the trend observed in field stars. As proposed by \citet{2017A&A...600A..70A} and \citet{2021MNRAS.503.3279S}, this effect could be due to a survival bias for open clusters, as high metallicity clusters born in the inner disc are more likely to survive if they migrate outwards to Galactic regions where the potentials of the spiral arm and bar are weaker \citep[see also][]{1987A&A...188...35L}{}{}. Low metallicity clusters from the outer disc which migrate towards the Galactic centre are more likely to be disrupted and are, therefore, missing in the older population, which shows a steeper gradient as a result. We show the clusters used in the work of \citet{2022AJ....164...85M} in Figure \ref{fig:MCM} because they use the same chemical abundances as employed in this work. The metallicity offset can be clearly seen in the $\qty[]{6}{\giga\year} \leq \tau < \qty[]{10}{\giga\year}$ bin.

\subsubsection{Models}
Figure \ref{fig:MCM} also shows a comparison to the chemo-dynamical simulation of \citet[][hereafter MCM]{2013A&A...558A...9M, 2014A&A...572A..92M} and the underlying chemical evolution model \citep{2009IAUS..254..191C}. There is a great diversity of chemical evolution models and simulations \citep[e.g.][]{2009MNRAS.396..203S, 2009MNRAS.399.1145S, 2015A&A...580A.126K, 2015A&A...580A.127K, 2015MNRAS.447.4018G, 2017MNRAS.472.3637G, 2018MNRAS.481.2570G, 2020MNRAS.496...80V, 2021A&A...647A..73S} and we choose the MCM simulation to facilitate easy comparison with the works of \citet{2017A&A...600A..70A}, \citet{2017MNRAS.470.4363C, 2019MNRAS.490.1821C} and \citet{2022AJ....164...85M}. The model was computed in instantaneously mixed galactocentric annuli \qty[]{2}{\kilo\parsec} wide and for this plot, we interpolate between those annuli and transform from $R_\mathrm{Gal}$ to $R_\mathrm{g}$. The metallicity is scaled to reproduce the solar value at approximately the time and location of the Sun's birth ($\approx \qty[]{4}{\giga\year}$ ago and $R_\mathrm{Gal} \approx \qty[]{6}{\kilo\parsec}$; \citealt{2013A&A...558A...9M}). To allow a qualitative comparison to our results, the particles from the MCM are restricted to those with $Z_\mathrm{Gal} < \qty[]{1}{\kilo\parsec}$, since the scale height of the simulated disc is larger than the MW's thin disc.

\begin{figure*}
	\includegraphics[width=0.8\linewidth]{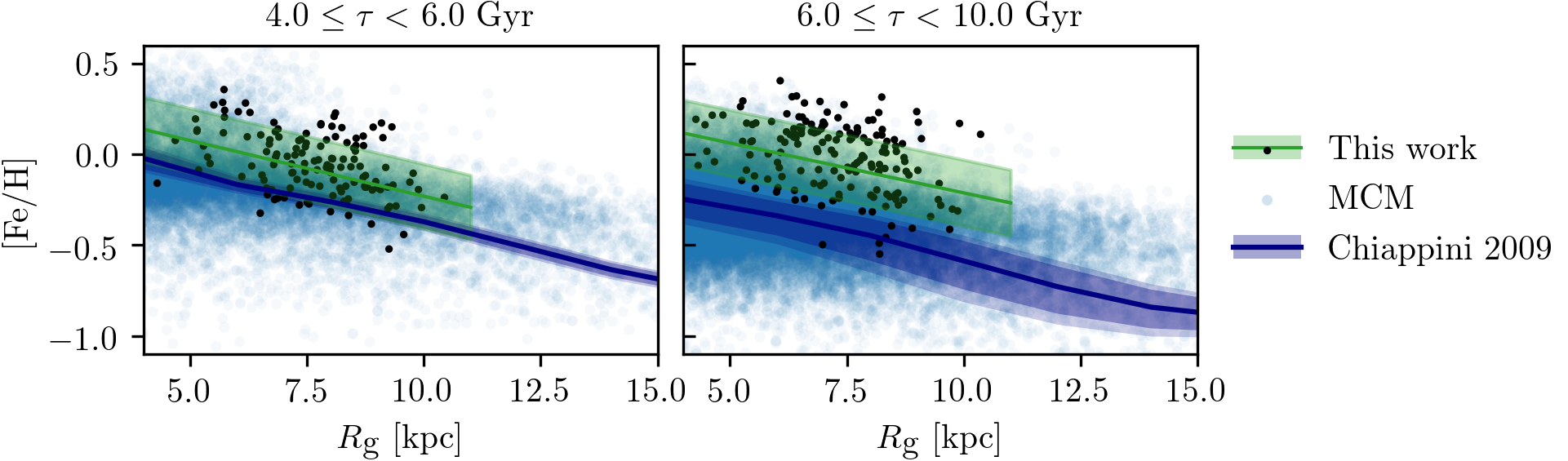}
    \caption{Same as Figure \ref{fig:MCM} for two age bins with the thin disc sample shown in black points, binned according to the median age of each star.}
    \label{fig:MCM_old}
\end{figure*}

Our results are reasonably consistent with the chemical evolution model and MCM simulation at ages below $\approx \qty[]{4}{\giga\year}$. In the older populations, the simulated particles show a more rapid flattening of the gradient, and an offset towards lower metallicities in the model becomes more apparent. This is likely due to the initial conditions of the models, as the thin disc is not pre-enriched in metals, giving a very metal-poor result at ages close to the start of disc formation. Our results, which are not subject to the metallicity bias which affects the older OCs, demonstrate that this very low metallicity is not a good representation of the thin disc's early state, and this effect is shown more clearly in comparison to the individual stars in our sample in Figure \ref{fig:MCM_old}.

\section{Conclusions}
\label{sec:Conclusions}

We have used the wide spatial and temporal coverage of the K2-APOGEE sample to study the evolution of the Milky Way's radial metallicity gradient in the range $\qty[]{4}{\kilo\parsec} < R_\mathrm{g} < \qty[]{11}{\kilo\parsec}$ across the whole history of the thin disc. We find a gradient which flattens monotonically from $\approx \qty[per-mode=symbol]{-0.07}{\dex\per\kilo\parsec}$ in stars of $\tau \approx \qty[]{0.7}{\giga\year}$ to $\approx \qty[]{-0.04}{\dex\per\kilo\parsec}$ at $\tau \approx \qty[]{12.6}{\giga\year}$. The intrinsic spread around this gradient, which is primarily caused by churning, increases with the age of the population until $\tau \approx \qty[]{10}{\giga\year}$. In older stars, the spread is slightly reduced, which we attribute to this population forming before the thin disc when star formation in the MW was very centrally concentrated. We also find that the metallicity of the solar neighbourhood has increased slightly over the last \qty[]{5}{\giga\year} to $\approx \qty[]{-0.04}{\dex}{}$, and that it is sub-solar throughout the history of the disc, indicating that the Sun migrated out from its birth radius to current position.

We have compared our results to other studies of field stars and explained the differences in our results in terms of the more robust selection of thin disc members and choice of guiding radius as the radial coordinate used in this work. We have also compared to recent studies of open clusters and have shown that older clusters appear to be subject to a survival bias, favouring those of higher metallicity. Finally, we have provided a qualitative comparison to the chemical evolution model of \citet{2009IAUS..254..191C} and the chemo-dynamical simulation of \citet{2013A&A...558A...9M, 2014A&A...572A..92M}. The most striking difference to our results is their much lower metallicity at ages above $\approx \qty[]{6}{\giga\year}$, which we attribute to the assumptions about the initial conditions in the model.  Our results will facilitate new tests of the initial conditions and physics in models and simulations of MW evolution.

In the next decade the ESA PLAnetary Transits and Oscillations of stars (PLATO) mission \citep{2014ExA....38..249R} can potentially provide ages with 10\% precision \citep{2017AN....338..644M} for a large sample of red giants. These data, combined with astrometry and spectroscopy from ongoing and planned surveys will allow the early evolution of the MW discs to be probed in detail and give a more precise measurement of the evolution of the metallicity gradient. These insights will throw new light on the open questions concerning the origin, strength and importance of radial migration in the Galactic archaeology fossil record.

\section*{Acknowledgements}


EW thanks Bill Chaplin and Daisuke Kawata for useful discussions and the anonymous reviewer for their helpful comments.

EW, AM, GC and VG acknowledge support from the ERC Consolidator Grant funding scheme (project ASTEROCHRONOMETRY, G.A. n. 772293 \url{http://www.asterochronometry.eu}).
CC acknowleges Fundacion Jesus Serra for its great support during her visit to IAC, Spain, during which part of this work was written.
AJL acknowledges the support of the Science and Technology Facilities Council.
This work was partially funded by the Spanish MICIN/AEI/10.13039/501100011033 and by the "ERDF A way of making Europe" funds by the European Union through grant RTI2018-095076-B-C21 and PID2021-122842OB-C21, and the Institute of Cosmos Sciences University of Barcelona (ICCUB, Unidad de Excelencia ’Mar\'{\i}a de Maeztu’) through grant CEX2019-000918-M. FA acknowledges financial support from European Union NextGenerationEU/PRTR and MCIN/AEI/10.13039/501100011033 through grant RYC2021-031638-I.

This paper includes data collected by the \textit{Kepler} mission and obtained from the MAST data archive at the Space Telescope Science Institute (STScI). Funding for the \textit{Kepler} mission is provided by the NASA Science Mission Directorate. STScI is operated by the Association of Universities for Research in Astronomy, Inc., under NASA contract NAS 5–26555.

This work has made use of data from the European Space Agency (ESA) mission {\it Gaia} (\url{https://www.cosmos.esa.int/gaia}), processed by the {\it Gaia} Data Processing and Analysis Consortium (DPAC, \url{https://www.cosmos.esa.int/web/gaia/dpac/consortium}). Funding for the DPAC has been provided by national institutions, in particular the institutions participating in the {\it Gaia} Multilateral Agreement.

Funding for the Sloan Digital Sky Survey IV has been provided by the Alfred P. Sloan Foundation, the U.S. Department of Energy Office of Science, and the Participating Institutions. 
SDSS-IV acknowledges support and resources from the Center for High Performance Computing  at the University of Utah. The SDSS website is \url{www.sdss4.org}.
SDSS-IV is managed by the Astrophysical Research Consortium for the Participating Institutions of the SDSS Collaboration including the Brazilian Participation Group, the Carnegie Institution for Science, Carnegie Mellon University, Center for Astrophysics | Harvard \& Smithsonian, the Chilean Participation Group, the French Participation Group, Instituto de Astrof\'isica de Canarias, The Johns Hopkins University, Kavli Institute for the Physics and Mathematics of the Universe (IPMU) / University of Tokyo, the Korean Participation Group, Lawrence Berkeley National Laboratory, Leibniz Institut f\"ur Astrophysik Potsdam (AIP),  Max-Planck-Institut f\"ur Astronomie (MPIA Heidelberg), Max-Planck-Institut f\"ur Astrophysik (MPA Garching), Max-Planck-Institut f\"ur Extraterrestrische Physik (MPE), National Astronomical Observatories of China, New Mexico State University, New York University, University of Notre Dame, Observat\'ario Nacional / MCTI, The Ohio State University, Pennsylvania State University, Shanghai Astronomical Observatory, United Kingdom Participation Group, Universidad Nacional Aut\'onoma de M\'exico, University of Arizona, University of Colorado Boulder, University of Oxford, University of Portsmouth, University of Utah, University of Virginia, University of Washington, University of 
Wisconsin, Vanderbilt University, and Yale University.

\textit{Software:} \texttt{ArviZ} \citep{2019JOSS....4.1143K}, \texttt{astropy} \citep{2013A&A...558A..33A, 2018AJ....156..123A, 2022ApJ...935..167A}, \texttt{corner} \citep{2016JOSS....1...24F}, \texttt{emcee} \citep{2013PASP..125..306F}, \texttt{galpy} \citep{2015ApJS..216...29B}, \texttt{JAX} \citep{jax2018github}, \texttt{Matplotlib} \citep{2007CSE.....9...90H}, \texttt{NumPy} \citep{2020Natur.585..357H}, \texttt{NumPyro} \citep{bingham2019pyro, 2019arXiv191211554P}, \texttt{pandas} \citep{reback2020pandas, mckinney-proc-scipy-2010}, \texttt{SciPy} \citep{2020NatMe..17..261V} and \texttt{TOPCAT} \citep{2005ASPC..347...29T}. We also acknowledge the use of the \texttt{Daft} package to produce figures in this work (\url{https://docs.daft-pgm.org/en/latest/}).




\section*{Data Availability}

The data underlying this article will be shared on reasonable request to the corresponding author.



\bibliographystyle{mnras}
\bibliography{refs} 

\appendix

\section{Effect of [F\MakeLowercase{e}/H] error inflation}
\label{app:feherr}

\begin{figure}
	\includegraphics[width=\linewidth]{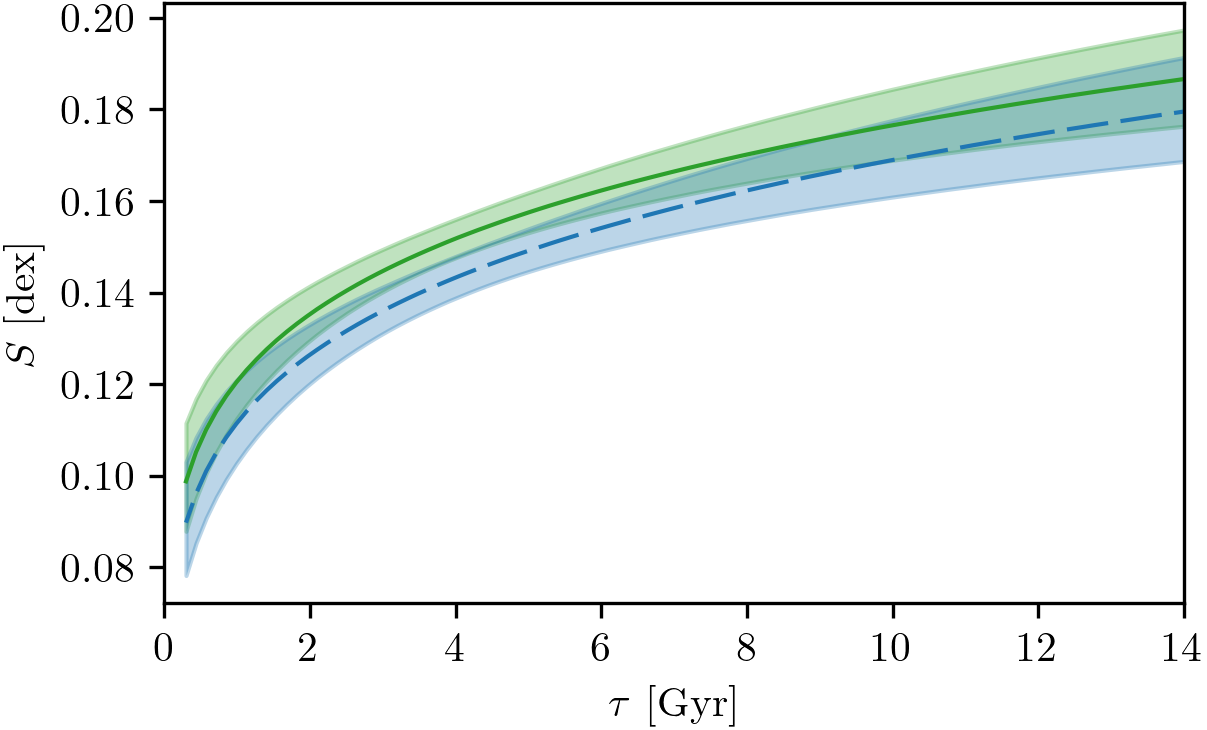}
    \caption{The [Fe/H] component of the intrinsic spread constrained by our HBM using the published [Fe/H] uncertainties (green) or inflating them as in  Willett et al. (in preparation; dashed blue).}
    \label{fig:HBM_feherr}
\end{figure}

As described in  Willett et al. (in preparation), during the PARAM runs to constrain the stellar parameters used in this work, the uncertainty on [Fe/H] from APOGEE DR17 was inflated to \qty[]{0.05}{\dex} for all stars with a reported uncertainty below this value. In our subsequent analysis of the metallicity gradient, we use the published values of uncertainty, but here we show the results obtained from the HBM if we follow the same inflation procedure. The evolution and uncertainties for the gradient and offset are almost unchanged from our original analysis and are not shown here, and Figure \ref{fig:HBM_feherr} shows that, though the intrinsic spread is slightly decreased, it is still fully consistent within uncertainties.

\section{Sensitivity to the $Z_\mathrm{\MakeLowercase{max}}$ cut}
\label{app:zmax}

\begin{figure}
	\includegraphics[width=\linewidth]{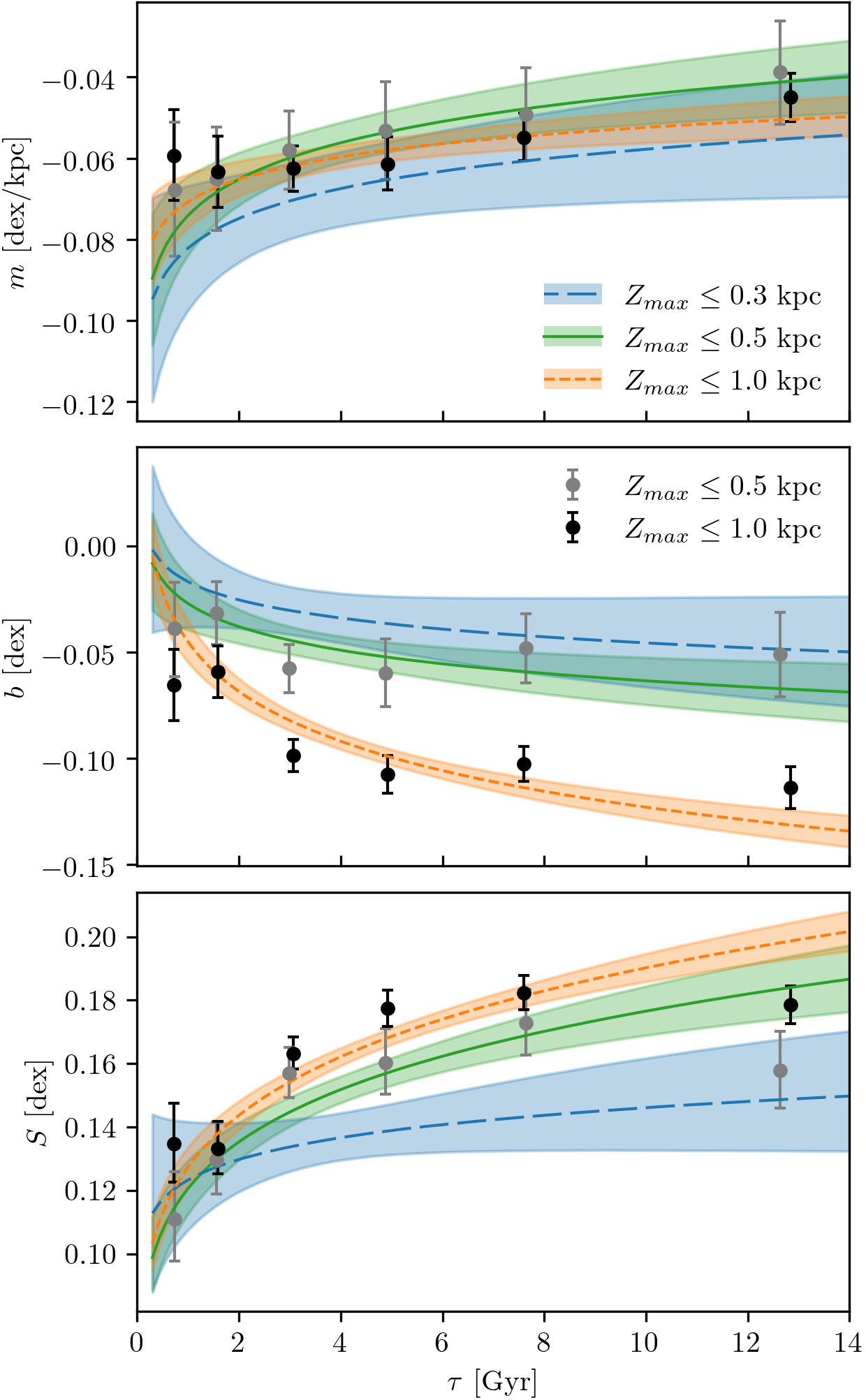}
    \caption{Evolution of the fit parameters as a function of stellar age. The grey points and green line and filled region show the best-fitting value and 68\% credible interval in each bin from the MCMC analysis (plotted against the median age of the population in the bin) and the result and 68\% credible interval from the HBM from the original analysis. The black points and short-dashed orange line and filled region show the same but for a sample including stars with a wider range of $Z_\mathrm{max}$. The long-dashed blue line and filled region show the HBM result for a narrower $Z_\mathrm{max}$ range.}
    \label{fig:HBM_Zcuts}
\end{figure}

\begin{figure}
	\includegraphics[width=\linewidth]{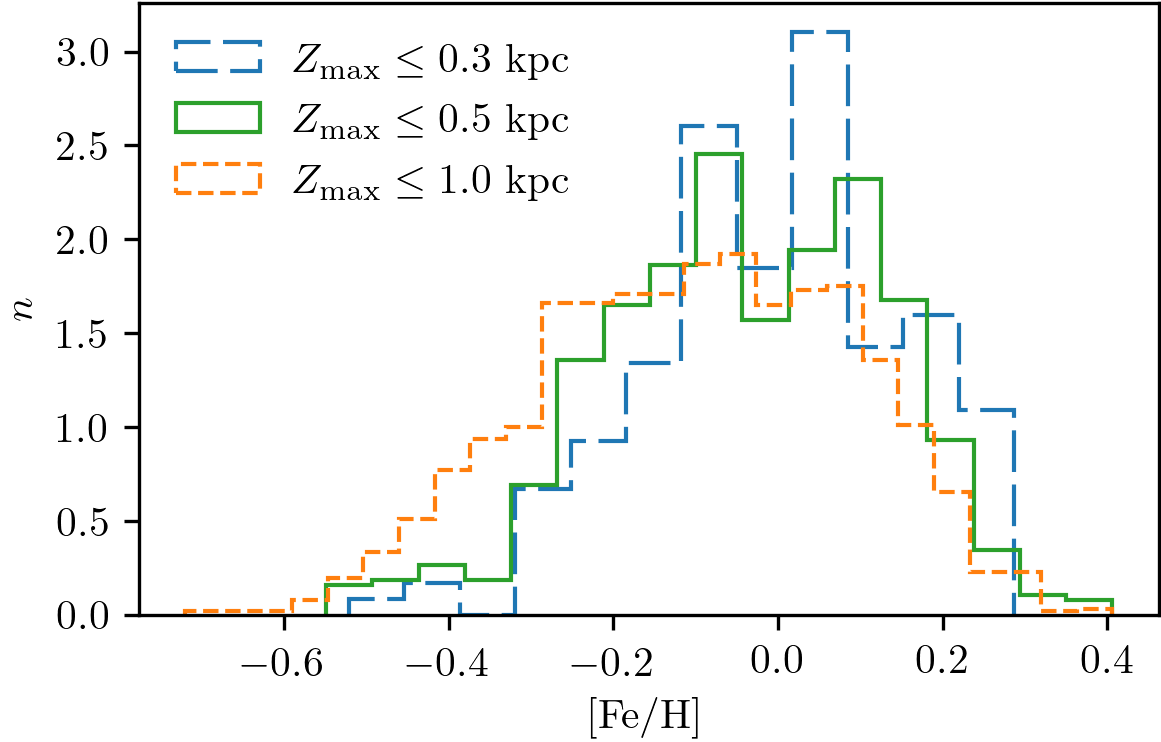}
    \caption{[Fe/H] distributions for samples including stars with a wider (short-dashed orange) or narrower (long-dashed blue) range of $Z_\mathrm{max}$ than the original analysis (green). The area under each histogram integrates to one.}
    \label{fig:FeH_dists}
\end{figure}

In this section, we show the sensitivity of our results to the cut in $Z_\mathrm{max}$ used to identify thin disc members. Figure \ref{fig:HBM_Zcuts} summarises the results of the MCMC and HBM analysis under different conditions. The gradient (\textit{top panel}) recovered with a wider ($Z_\mathrm{max} \leq \qty[]{1.0}{\kilo\parsec}$) or narrower ($Z_\mathrm{max} \leq \qty[]{0.3}{\kilo\parsec}$) range of $Z_\mathrm{max}$ is consistent with our original analysis within the uncertainties. Decreasing the $Z_\mathrm{max}$ range does not significantly change the [Fe/H] offset at \qty[]{8}{\kilo\parsec} (\textit{middle panel}), but the result from the wider cut shows significantly lower metallicity. This is consistent with the wider range of metallicities in this sample (Figure \ref{fig:FeH_dists}). The intrinsic spread (\textit{bottom panel}) of the three samples is more sensitive to the $Z_\mathrm{max}$ cut, but the relationship between the MCMC and HBM result is the same in a given sample.

\section{Effect of $R_\mathrm{\MakeLowercase{g}}$ or $R_\mathrm{G\MakeLowercase{al}}$}
\label{app:RGal}

In this section, we provide further detail on the sensitivity of our results to the radial coordinate and range used in the analysis.

\begin{figure*}
	\includegraphics[width=0.9\linewidth]{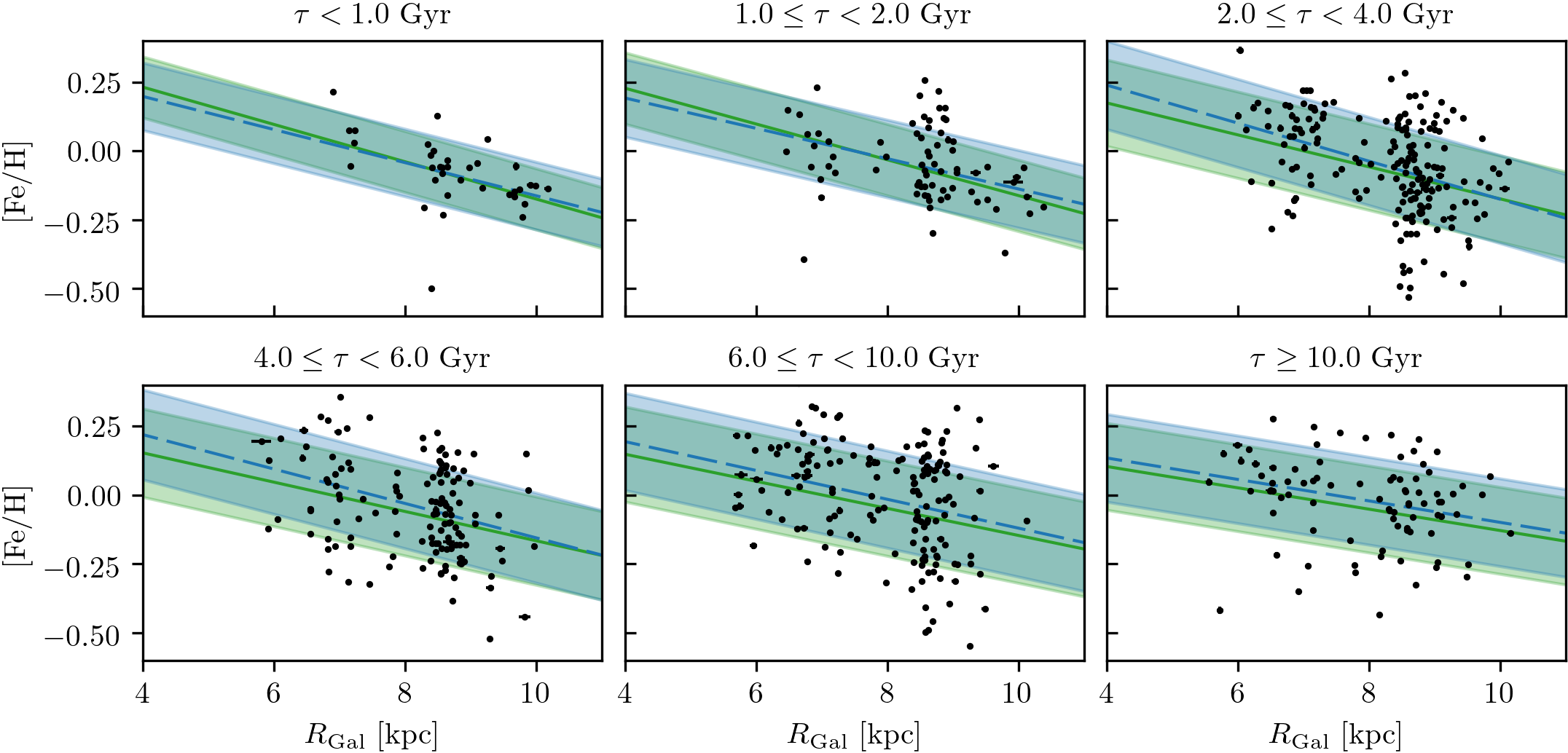}
    \caption{[Fe/H] vs. $R_\mathrm{Gal}$ in six bins of age. The thin disc sample is shown in black points, binned according to the median age of each star. The dashed blue line and filled region show the results from the MCMC analysis using $R_\mathrm{Gal}$. For comparison, the green line and filled region show the results using $R_\mathrm{g}$ (plotted against $R_\mathrm{g}$).}
    \label{fig:bins_Rgal}
\end{figure*}

\begin{figure}
	\includegraphics[width=\linewidth]{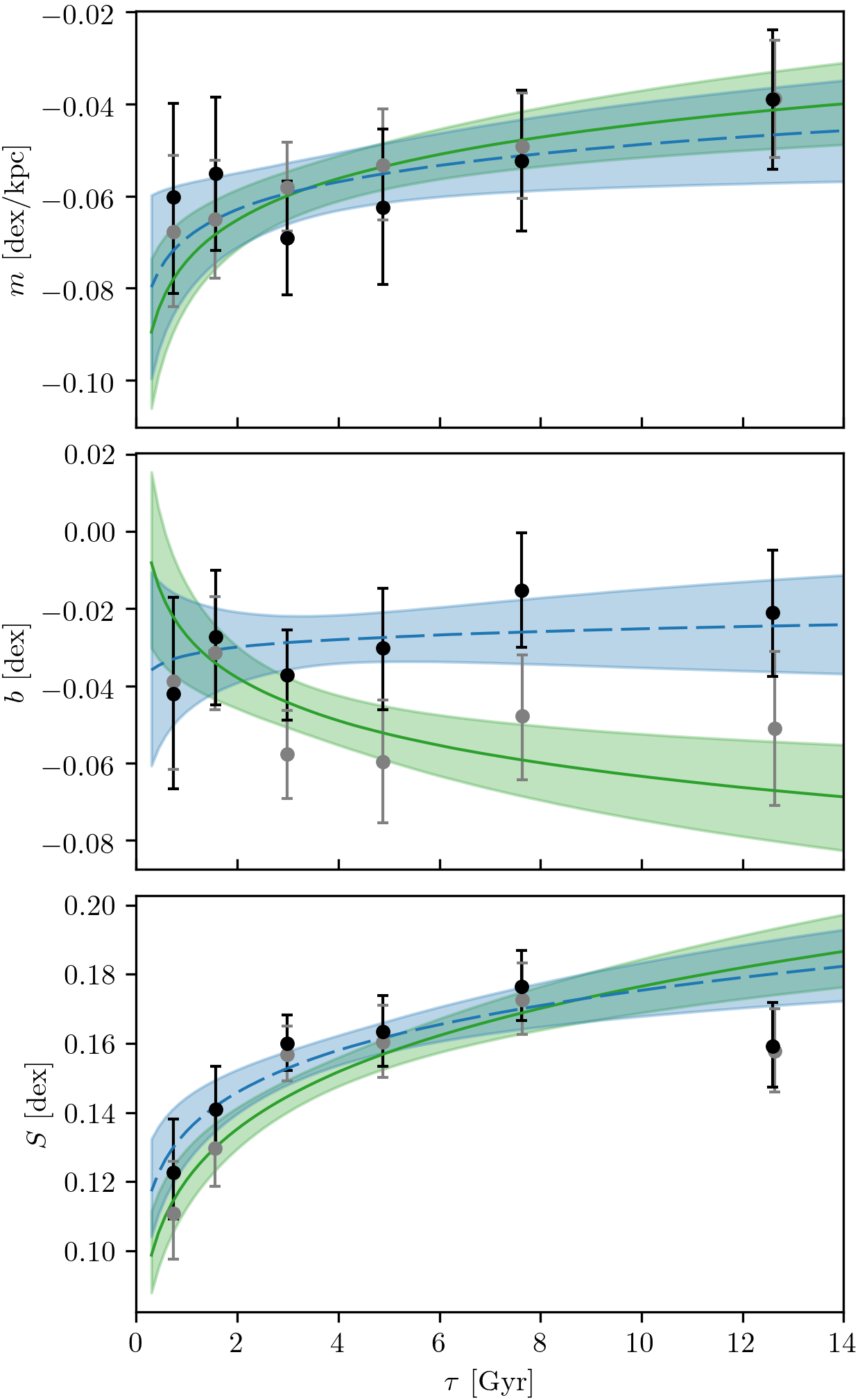}
    \caption{Evolution of the fit parameters as a function of stellar age. The grey points and green line and filled region show the best-fitting value and 68\% credible interval in each bin from the MCMC analysis (plotted against the median age of the population in the bin) and the result and 68\% credible interval from the HBM from the original analysis using $R_\mathrm{g}$. The black points and dashed blue line and filled region show the same using $R_\mathrm{Gal}$.}
    \label{fig:HBM_binsRgal}
\end{figure}

First, we show the results of our analysis using $R_\mathrm{Gal}$ rather than $R_\mathrm{g}$. Figure \ref{fig:bins_Rgal} shows the results in bins of age, where the uneven radial distribution caused by the K2 field placement can be clearly seen. Figure \ref{fig:HBM_binsRgal} summarises the results from the MCMC and HBM analysis, where in general all parameters agree within their uncertainties. The exception is the HBM result for the [Fe/H] offset at \qty[]{8}{\kilo\parsec}, which is likely a result of the populations clumping up in $R_\mathrm{Gal}$.

We also test the sensitivity of our result to stars with $R_\mathrm{g} < R_\mathrm{Gal,\,min}$ (see Section \ref{ssec:sel_eff} for details). Excluding these stars does result in a slightly steeper gradient at $\tau > \qty[]{1}{\giga\year}$, but the two results are consistent within uncertainties. The offset and intrinsic spread are almost unchanged in the new samples and are not shown here. This indicates that our results are not biased by the choice of $R_\mathrm{g}$ as the radial coordinate in our analysis.

\begin{figure}
	\includegraphics[width=\linewidth]{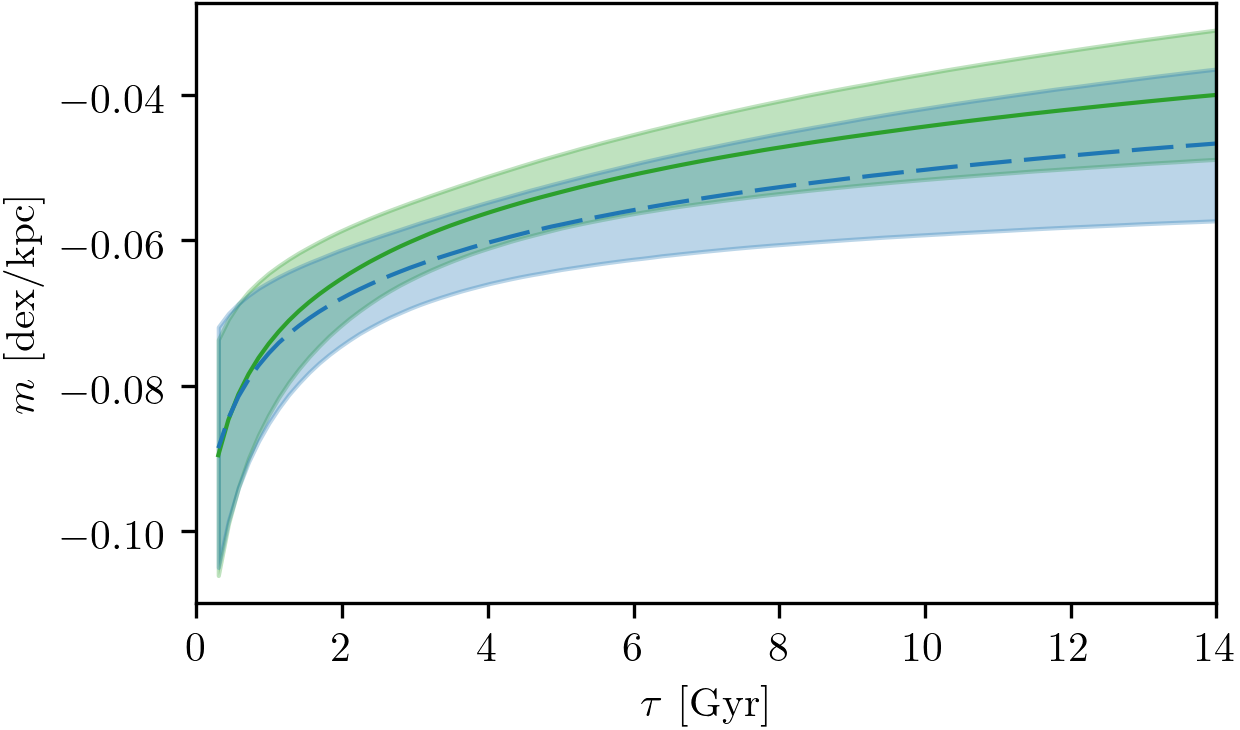}
    \caption{Evolution of the gradient with age for the full sample (green) and excluding stars with $R_\mathrm{g} < R_\mathrm{Gal,\,min}$ (dashed blue).}
    \label{fig:HBM_Rg55}
\end{figure}

\bsp	
\label{lastpage}
\end{document}